\documentclass[useAMS,usenatbib,twocolumn]{mnras}
\usepackage[dvipdf]{graphicx}
\usepackage{txfonts}
\usepackage{natbib}
\usepackage{bm}

\def\apj{ApJ}
\def\apjl{ApJL}

\def\aap{A\&A}

\def\mnras{MNRAS}

\def\beq#1{\begin{equation}\label{#1}}
\def\eeq{\end{equation}}
\def\beqa#1{\begin{eqnarray}\label{#1}}
\def\eeqa{\end{eqnarray}}
\def\eq#1{eq.~(\ref{#1})}

\def\comment#1{\relax}

\title[Three-dimensional modelling of accretion columns]
{Three-dimensional modelling of accretion columns: spatial asymmetry and self-consistent simulations
}
\author[M. I. Gornostaev]{M. I. Gornostaev$^{1,2}$\thanks{E-mail: mgornost@gmail.com}\\
$^{1}$ Sternberg Astronomical Institute, Lomonosov Moscow State University, Universitetskij pr., 13, Moscow 119234, Russia\\
$^{2}$ Faculty of Physics, Lomonosov Moscow State University, Leninskie Gory, 1, Moscow 119991, Russia
}

\begin{document}

\date{Accepted XXX. Received YYY; in original form ZZZ}
\pagerange{\pageref{firstpage}--\pageref{lastpage}} \pubyear{2020}

\maketitle

\label{firstpage}

\begin{abstract}
The paper presents the results of three-dimensional (3D) modelling of the structure and the emission of
 accretion columns formed above the surface of accreting strongly magnetized neutron stars
under the circumstances when a pressure of the photons generated in the column base is enough to determine
the dynamics of the plasma flow. On the foundation of numerical radiation hydrodynamic simulations, several 3D models of
accretion column  are constructed. The first group of the models contains spatially 3D columns. The corresponding
calculations lead to the distributions of the radiation flux over the sidewalls of the columns which are not characterized by axial symmetry.
The second group includes the self-consistent modelling of spectral radiative transfer and two-dimensional spatial structure of
the column, with both thermal and bulk Comptonization taken into account. The changes in the structure of the column and the shape of
X-ray continuum are investigated depending on physical parameters of the model.

\end{abstract}


\section{Introduction}
\label{s:intro}
The location of the emitting regions of accreting strongly magnetized neutron stars is controlled by the magnetic field, which
is often assumed to be dipole in solving the problems. If the axis of a dipole does not pass
through the neutron star centre, the modelling  of an emitting structure cannot be performed
 under the assumption of  axial symmetry of the problem, which is three-dimensional (3D) in this case.
The same is rightful when the accretion channel near the neutron star surface has a transverse section of comparatively
complicated form. These are the reasons for considering 3D models of radiation-dominated
accretion columns, whose two-dimensional (2D) structures  were calculated earlier (\citealt{1973NPhS..246....1D},
\citealt{1981A&A....93..255W}, \citealt{2015MNRAS.452.1601P}). The  numerical solutions presented below give the impression
of the quantitative deviation of the  shock form  from the 2D axially symmetric one  and lead to the
2D distributions of radiative flux over the column sidewall. The consequences of the
presented consideration can be useful for the investigation of properties of spatially one-dimensional (1D)
solutions (\citealt{2007ApJ...654..435B}, \citealt{2012A&A...538A..67F},
\citealt{2016A&A...591A..29F}, \citealt*{2017ApJ...835..129W, 2017ApJ...835..130W}), as well as
for the modelling and interpretation of the observed pulse profiles of X-ray pulsars emission.

Another kind of 3D problems is related with a transition to  simultaneous modelling of
the dynamics of the matter and radiative transfer.  The solution procedure offered does not
involve the frequency-integrated energy equation. One can notice this immediately comparing the
results of Section \ref{sec:3d}, where spatially 3D computations are described,  with the results
of Section \ref{sec:3dsp}, devoted to the modelling of spatially 2D radiation-dominated shocks with
simultaneous calculation of spectral radiation energy density at each point of the column.
 Thus, hereafter a self-consistent solution will be understood as a set of functions that are a solution
of one and the same system of equations and describe the velocity of the plasma flow, the electron temperature distribution and the
radiation spectrum within a column.

Both approaches are based, first of all, on the assumption that the matter flow satisfies the hydrodynamic limit.
The Compton scattering is considered as the main process
of the energy exchange between the plasma and the radiation field.
The  basic equations under consideration describe the propagation of the radiation not
only along the magnetic field but also in transverse directions, governing the mound-like structure of the
shock and determining the corresponding distributions of the
radiation flux over the column sidewall.

Section \ref{sec:concl} contains some remarks and conclusions.

\newpage

\section{Three-dimensional radiation-dominated shocks}
\label{sec:3d}
\subsection{The relations between main equations}
Under the circumstances of the radiation-dominated accretion columns, the structure formed above the magnetic
poles has a significant optical depth to the Thomson scattering, and the radiative pressure is much
higher here than the gas one. These things allow to believe that the diffusion approximation is applicable to
describe the radiative transfer, and that the pressure tensor is isotropic.
In all frames considered below,  the neutron star is at rest.
Then, the radiation flux within the column can be written as
\beq{e:flux}
\bm F = -D\nabla u +\frac{4}{3}u{\bm \varv},
\eeq
and the momentum equation has the form
\beq{e:momentum}
n_{\rm e}m_{\rm p}({\bm \varv}\cdot\nabla){\bm \varv}=-\frac{1}{3}\nabla u,
\eeq
where $u$ denotes the frequency-integrated radiation energy density, $\bm \varv$
is the bulk velocity of the matter, $n_{\rm e}$ is the electron number density, $m_{\rm p}$
is the proton mass, and $D$ is the diffusion coefficient.
The gravitation and the gas pressure are neglected in the equation (\ref{e:momentum}).
The considered physical situation, satisfying the limit
$p_{\rm g}/p_{\rm r}\ll 1$, where  $p_{\rm g}$ and $p_{\rm r}$ denote the gas pressure and the radiation one, respectively,
allows one not only to set $p=p_{\rm r}=u/3$ in \eq{e:momentum}, but also
not to consider further the equations of state for both electron and ion gas components.

Let us assume that magnetic field is uniform inside  the computational domains.
The velocity vector will be supposed to be collinear to the magnetic field vector, which corresponds to situation when the matter is
confined in the transverse directions by a magnetic field whose pressure far exceeds any other within the problem.
The stationary continuity equation then reads as
\beq{e:cont}
n_{\rm e}m_{\rm p}\varv=S,
\eeq
where $\varv=|\bm \varv|$, $S=\dot M/A$, with $\dot M$ is the mass accretion
rate per one neutron star magnetic pole and $A$ is a column transverse section area.
The integration of \eq{e:momentum} yields
\beq{e:r-density}
\varv=\varv_0-\frac{u}{3S}.
\eeq
Here, $\varv_0$ is the absolute value of the velocity at a height, where one can set $u=0$ (far above the neutron star surface).

The stationary equation for the frequency-integrated radiation energy density \citep{1981MNRAS.194.1033B} has the form
\beq{e:kinint}
\nabla \cdot\left(D\nabla u\right) - \bm \varv \nabla u - \frac{4}{3}u \nabla\!\cdot\!\bm \varv +
\frac{u n_{\rm e}\sigma_{\rm T}}{m_{\rm e}c}\left(4kT_{\rm e}-\langle\epsilon\rangle\right)
=0,
\eeq
where $k$ is the Boltzmann constant, $m_{\rm e}$ is the electron mass,
 $T_{\rm e}$ is the electron temperature, the scattering cross-section is set to be equal to the Thomson value, $\sigma_{\rm T}$,
 and the mean (energy-weighted) value $\langle\epsilon\rangle$ of the photon energy,  $\epsilon$, is related with
angle-averaged photon occupation number  $n$ as follows,
\beq{}
\langle\epsilon\rangle=\frac{\int  \epsilon^4 n d\epsilon}{\int  \epsilon^3 n d\epsilon}.
\eeq
After substituting in \eq{e:kinint} the temperature \citep{1970JETPL..11...35Z}
\beq{e:Te}
T_{\rm e}=\frac{1}{4k}\frac{\int \epsilon^4 n d\epsilon}{\int \epsilon^3 n d\epsilon}
\eeq
 which is equal to the electron one in the approximation of the local Compton equilibrium,
the term corresponding to the energy-integrated Kompaneets operator \citep{1956Kompaneets} becomes equal to zero, so that
\beq{e:endensint}
\nabla \cdot\left(D\nabla u\right) - \frac{4}{3} u \nabla\!\cdot\!\bm \varv - \bm \varv\cdot\nabla u=0.
\eeq
The term $n^2$ is neglected in expression (\ref{e:Te}).
From \eq{e:endensint} it follows that
\beq{}
\nabla \cdot \left(-D\nabla u+\frac{4}{3}u\bm \varv\right)=\frac{1}{3}\bm \varv\cdot\nabla u,\nonumber
\eeq
and thus the energy equation can be written as
\beq{e:energy1}
\nabla \cdot \bm F
= -n_{\rm e} m_{\rm p} \bm \varv\cdot\nabla\left(\frac{\varv^2}{2}\right).
\eeq

Previous (two-dimensional) calculations are based on the solution of the energy equation of type of equation (\ref{e:energy1})
making use of the velocity value determined by momentum equation and taking into account
the continuity equation (\citealt{1973NPhS..246....1D} and references to this work).
The denoted relation between equations (\ref{e:kinint}) and (\ref{e:energy1}) plays a determinative role for constructing the solutions,
presented in Section \ref{sec:3dsp}.

\subsection{Geometry}

It is reasonable to carry out spatially 3D calculations in Cartesian coordinates.
Let us introduce them in such a way that the axis $z$ will have the direction opposite to the velocity of the flow.
Let the idealized neutron star surface being approximated by the plane inside the computational domains
 intersect the $xy$ plane, and let $\alpha$ denote the angle between these planes  ($\alpha<\upi/2$).
Meanwhile, let the neutron star surface intersect the $z$ axis in the half-space $z>0$.

Two kinds of column geometry are considered numerically in the current section.

\textit{Cylindrical filled column.} Consider the column in the model of circle truncated cylinder with radius $r_0$,
the continuity equation includes the area $A=\upi r_0^2$ in this case.
Let the axis of the cylinder be superposed with the $z$ axis,
and the straight line originated by the intersection of the $xy$ plane and neutron star surface be the
tangent to the cylinder sidewall at the point $x=r_0$, $y=0$, $z=0$.

The value of $r_0$ being determined by the Alfven radius $r_{\rm A}$,
\beq{}
r_0=\beta r_{\rm ns}\sqrt{\frac{r_{\rm ns}}{r_{\rm A}}},
\eeq
where $r_{\rm ns}$ is the neutron star radius, is not defined strictly since the factor
$\beta\sim 1$ depends on the freezing depth of the accreted plasma in the magnetosphere and
the geometry of the accretion flow beyond the Alfven surface.

\textit{Unclosed hollow column.} On the base of  descriptions of possible accretion
column geometry (\citealt{1976MNRAS.175..395B}, \citealt{1984SSRv...38..325M}, \citealt{2015MNRAS.447.1847M}, \citealt{2015MNRAS.452.1601P}),
in the present work the modelling has been carried out also for another case.
This last corresponds to  the arch-like form  of the channel transverse section and
is realized in the frame of consideration of two circle cylinders with the axes parallel to the $z$ axis.
Namely, let the axis of cylinder of radius $r_0$ be superposed with the $z$ axis,
and the axis of cylinder of radius $r_0 + br_0$ intersect the $xy$ plane at the point
$x = s$, $y = 0$, with $br_0<s<(2+b)r_0$.
Then the directrices of cylinders lying in the same plane intersect at two points with the same abscissa $x^*$.
Consider the arcs lying on the same side of the chord passing through the intersection points in the case of $x^*<0$.
For the definiteness let us choose two longer arcs and consider the figure enclosed between them.
Thus, the column sidewalls are generated by the generatrices of cylinders crossing these arcs,
and the continuity equation (\ref{e:cont}) includes in considering case the area $A$ of specified figure.
It can be calculated as a difference between areas of two corresponding circle segments:
$A=A_2-A_1$, where $A_1=\frac{r_0^2}{2}(\psi_1-\sin\psi_1)$, $A_2=\frac{(r_0+br_0)^2}{2}(\psi_2-\sin\psi_2)$,
with $\psi_1=2\arccos\left(\frac{x^*}{r_0}\right)$,  $\psi_2=2\arccos\left(\frac{x^*-s}{r_0+br_0}\right)$.
The straight originated by the intersection of the $xy$ plane and neutron star surface will be the
tangent to column sidewall at the point $x=r_0+br_0$, $y=0$, $z=0$.

The modelling is feasible as well in the geometry of closed hollow columns (when $|s|\leq b r_0$, and $A=(2+b)b\upi r_0^2$).
The case of $s=0$ and $\alpha=0$ corresponds to 2D problem of ring-kind geometry, considered numerically by \cite{2015MNRAS.452.1601P}.
The numerical consideration of so-called spaghetti-like geometry \citep{1984SSRv...38..325M} can be reduced to the modelling
of a set of radiation-dominated columns while the approach described above is applicable.

\subsection{Computations}

Let us now describe the solution of the system of equations containing
(\ref{e:flux}), (\ref{e:cont}), (\ref{e:r-density}) and (\ref{e:energy1}),
and coming to the one partial differential equation of the elliptic type.

 The transition to the dimensionless variables is performed by the changes:
\beq{e:Q}
\tilde{x}=\frac{\sigma_{\rm T}  S}{m_{\rm p}c} x,~\tilde{y}=\frac{\sigma_{\rm T}  S}{m_{\rm p}c} y,~\tilde{z}=\frac{\sigma_{\rm T}S}{m_{\rm p}c} z,~Q=\frac{\varv^2}{\varv_0^2},
\eeq
where  $c$ is the speed of light,  and
$\varv_0$ being now specified equals
to the free-fall velocity at the upper boundary,
\beq{e:v0}
\varv_0=\sqrt{\frac{2GM_{\rm ns}}{r_{\rm ns}+z_0}},
\eeq
with $G$ is the gravitation constant,  $z_0$ is the coordinate of the upper boundary;
hereafter, it is set that $r_{\rm ns}=10^6~{\rm cm}$ and the neutron star mass $M_{\rm ns}=1.5M_\odot$.

  The question of the specific form of the diffusion
 coefficient was not considered above.
 Its definition is related with the method of
 the describing of scattering in the strong magnetic field used by \cite{1981A&A....93..255W}, \cite{2007ApJ...654..435B},
\cite{2017ApJ...835..129W}.
 This modification of the grey approximation is aimed at the effective accounting
 on the radiative transfer of the angular and frequency dependence of scattering cross-sections for ordinary and extraordinary modes.

The approximation is constructed  by analyzing the frequency dependence of non-relativistic scattering cross-sections
derived by \cite*{1971PhRvD...3.2303C}.
One can consider two narrow ranges of the angle $ \theta $ of the initial direction of the wavevector
with respect to the magnetic field vector. Let the condition $\epsilon \gtrsim h \nu_{\rm L}$ be satisfied,
where $ \nu_{\rm L}$ is the plasma frequency and $h$ is the Planck constant. For $\theta$ that are close to 0,
the cross-sections read as (\citealt{1971PhRvD...3.2303C}, \citealt{1974ApJ...190..141L})
\beq{e:CS1}
\sigma_j \simeq \sigma_{\rm T}\left (\frac{\epsilon ^ 2 }{(\epsilon + (- 1) ^ j \epsilon_B)^2} + \frac {1}{2} \sin ^ 2 \theta \right),
\eeq
where the cyclotron energy $\epsilon_B=h eB / (2\upi m_{\rm e} c)$,
$e$ is the elementary charge,
$j=1$ for the extraordinary mode, and $j=2$ for the ordinary one.
For  $\theta$ that are close to $\upi/2$,
\beqa{e:CS2}
\sigma_1 \simeq \sigma_{\rm T} \left (\frac{\epsilon ^ 2}{(\epsilon- \epsilon_B) ^ 2} + \cos ^ 2 \theta \right),\\\nonumber
\sigma_2 = \sigma _ {\rm T} \sin ^ 2 \theta.
\eeqa
It can be seen from these expressions that while, for $ \epsilon \ll \epsilon_B $,
$ \sigma_1 \simeq \sigma _ {\rm T} (\epsilon / \epsilon_B) ^ 2 $ in both directions, $ \sigma_2 \simeq \sigma _ {\rm T}$ in
the perpendicular direction and $ \sigma_2 \simeq \sigma _ {\rm T}(\epsilon / \epsilon_B) ^ 2 $ in the parallel one;
for $ \epsilon \gg \epsilon_B $, $ \sigma_j = \sigma _ {\rm T} $, $j = 1, 2$. From here one can see the anisotropic
character of the scattering in the highly magnetized plasma, the effective account of which is based thus on
the judicious parametrisation of the value of ratio $(\epsilon^*/ \epsilon_{B})^2$, where the
mean energy  $\epsilon^*<\epsilon_B$ \citep{1981A&A....93..255W} replaces its specific value $\epsilon$.
 Thus, the main components of the diffusion tensor, describing the propagation of photons across and along the magnetic field,
 are taken to be equal to
 \beq{e:Dperp}
 D^\perp=\frac{c}{3\sigma_{\rm T}n_{\rm e}}
  \eeq
  and
 \beq{e:Dpar}
 D^\|=\left(\frac{\epsilon_{B}}{\epsilon^*}\right)^2 D^\perp,
 \eeq
 respectively.
 In the calculations it is set  that $(\epsilon_{B}/\epsilon^*)^2=10$, excepting two cases described in the next section.

Notwithstanding the huge values of the scattering cross-sections on the cyclotron resonance,
the radiation pressure force distinguishes significantly from its values in the case of
non-magnetized atmosphere, as \cite{1984A&A...138..356G} showed, only in superficial
layers (the plane homogeneous  hot  magnetized atmosphere was being investigated).
It is possible to believe,  consequently, that the cyclotron photons are not able to influence
the dynamics of the flow dramatically.

The specific value of magnetic field strength is not involved explicitly in the scattering cross-sections under the current consideration.
Therefore, the values of $r_0$ and $D^\|$ will be varied independently.

\begin{figure}
  	\begin{center}
  \includegraphics[width=0.45\textwidth, trim=35mm 80mm 40mm 80mm, clip=true]{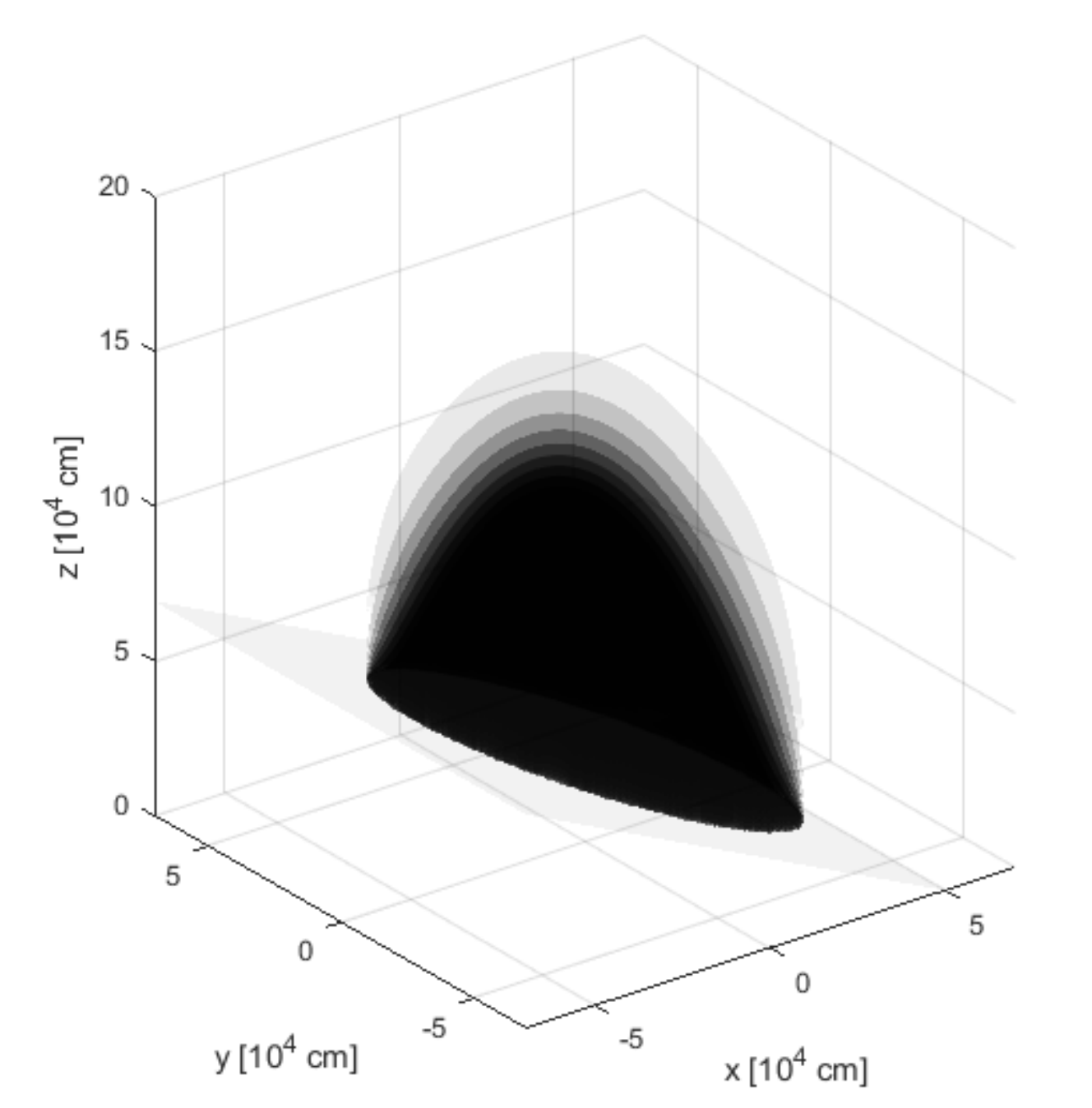}\\(a)\\
  \includegraphics[width=0.45\textwidth, trim=35mm 80mm 40mm 80mm, clip=true]{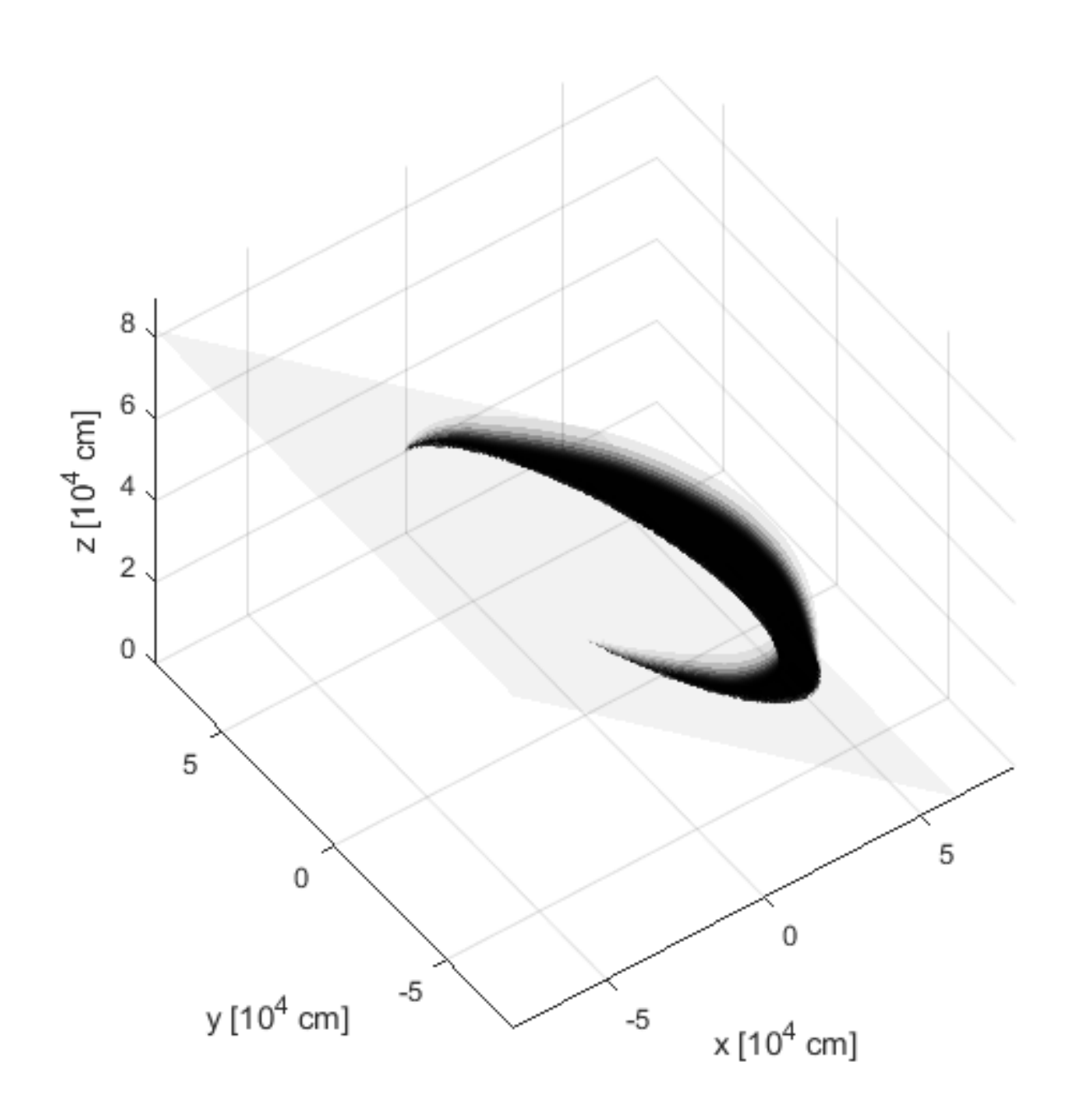}\\(b)
	\caption{Three-dimensional structure of the radiation-dominated column simulated in the geometry of the filled truncated cylinder (a)
and unclosed narrow circlet (b). The level surfaces of quantity $Q$ ($0.9, 0.8, ...$) are shown for $\alpha=30^\circ$,
the other parameters are specified in the main text. The neutron star surface is represented by the grey plane.
}
  \label{fig:3d}
	\end{center}
 \end{figure}

 Turning now to the equations (\ref{e:flux}), (\ref{e:cont}), (\ref{e:r-density}), and (\ref{e:energy1}), after the substitutions one can write in the dimensionless form the following equation:
 \beq{e:Qeq}
\frac{\partial^2 Q}{\partial \tilde{x}^2}+\frac{\partial^2 Q}{\partial \tilde{y}^2} + \frac{D_\|}{D_\perp}\frac{\partial^2 Q}{\partial \tilde{z}^2}-\frac{\partial}{\partial \tilde{z}}\left(8\sqrt{Q}-7Q\right)=0.
\eeq

The numerical solution of this equation is carried out by the time relaxation method,
which consists in searching the stationary solution of the problem for parabolic equation
including the same spatial operator as the original elliptic equation.
The difference schemes are explicit, and the rectangular equidistant meshes are used.
The second derivatives are approximated by central three-point finite-difference patterns,
and the first derivatives are approximated by central differences.
All the calculations described in the current and next sections are performed
using the programs written in {\sc C} language,
the computer based on Intel Core i9-9900KF CPU is exploited.

Since the main fraction of the kinetic energy of the matter transforms into the radiation energy in the shock,
near the star surface the appropriate condition is $Q=0$,
which leads to the infinite density of the matter at this  boundary
(the neutron star surface is impenetrable for the matter flow).
For numerical reasons,
it can be assumed that the radiation energy density at the bottom boundary is \hbox{slightly (1--3~\%)}
less than the maximal value $3S\varv_0$.
 At the upper boundary, it is set that $Q=1$. The value of $z_0$ is chosen enough
 to be affected on the solution for the shock structure.
The radiation leaves the column and emerges to the surrounding space freely,
so that in dependence on required accuracy, one can use the condition
\beq{e:boundSW1}
Q=1,
\eeq
or the condition  for the projection of the radiation flux
vector onto the direction of a normal unit vector ${\bm n}$ to the outer side of column surface,
\beq{e:boundSW}
F_n=\kappa cu.
\eeq
Here,  $\kappa={\rm const \lesssim 1}$, and $F_n=-D\frac{\partial u}{\partial n}$,
where $\frac{\partial u}{\partial n}$ is a directional derivative of $u$ in the ${\bm n}$ direction.
For computational reasons,  the condition (\ref{e:boundSW}) is realized at the side surfaces of the unclosed hollow column,
and the condition (\ref{e:boundSW}) with $\kappa=2/3$ is set  in the model of the filled column.
The numerical realization of boundary condition  (\ref{e:boundSW}) is performed making use of the forward or backward
(in dependence on coordinate quarter) three-point finite-difference patterns approximating the first derivatives.

\subsection{Results}

The examples of 3D structures of accretion columns obtained by solving equation (\ref{e:Qeq}) are shown in Fig.~\ref{fig:3d}.
The structure of the filled column is displayed in Fig.~\ref{fig:3d}a, where the set of surfaces
of equal quantity $Q$ is plotted. For the specific computation, the  following values of the parameters are used:
the mass accretion rate
$\dot M_{17}=\dot M/(10^{17}~{\rm g~ s^{-1}})=1$,
the inclination angle $\alpha=30^\circ$, and the column radius $r_0=5\times 10^4 {\rm~cm}$.

Fig. \ref{fig:3d}b shows the level surfaces of $Q$ obtained due to numerical simulations in unclosed hollow column geometry
for the same set of parameters, with $b=0.1$ and $s=0.15 r_0$.
The code provides, in principle, the arbitrary spatial orientation of the emitting region.

\begin{figure}
  	\begin{center}
  \includegraphics[width=0.35\textwidth]{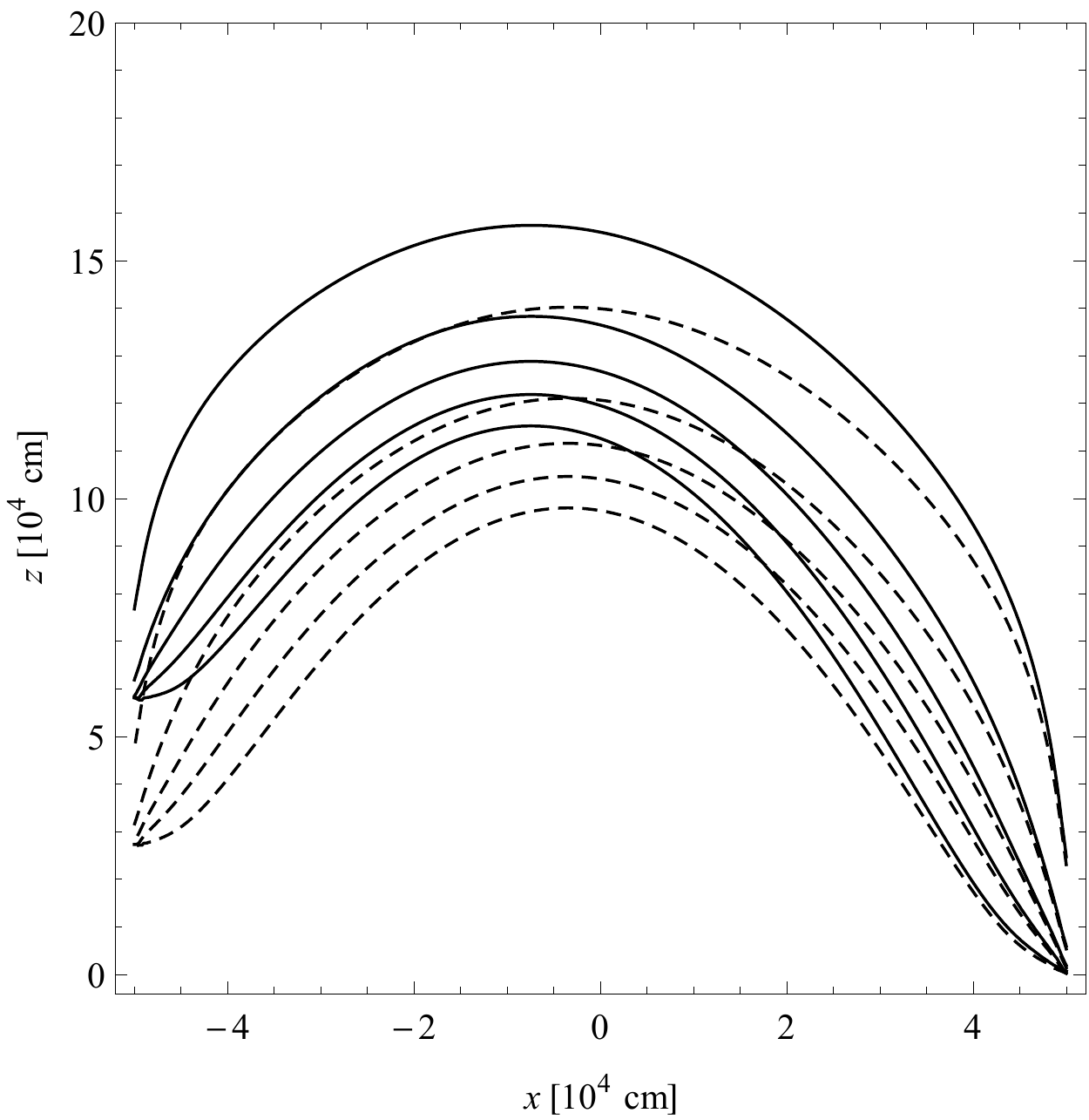}\\(a)\\\vspace{10pt}
  \includegraphics[width=0.35\textwidth]{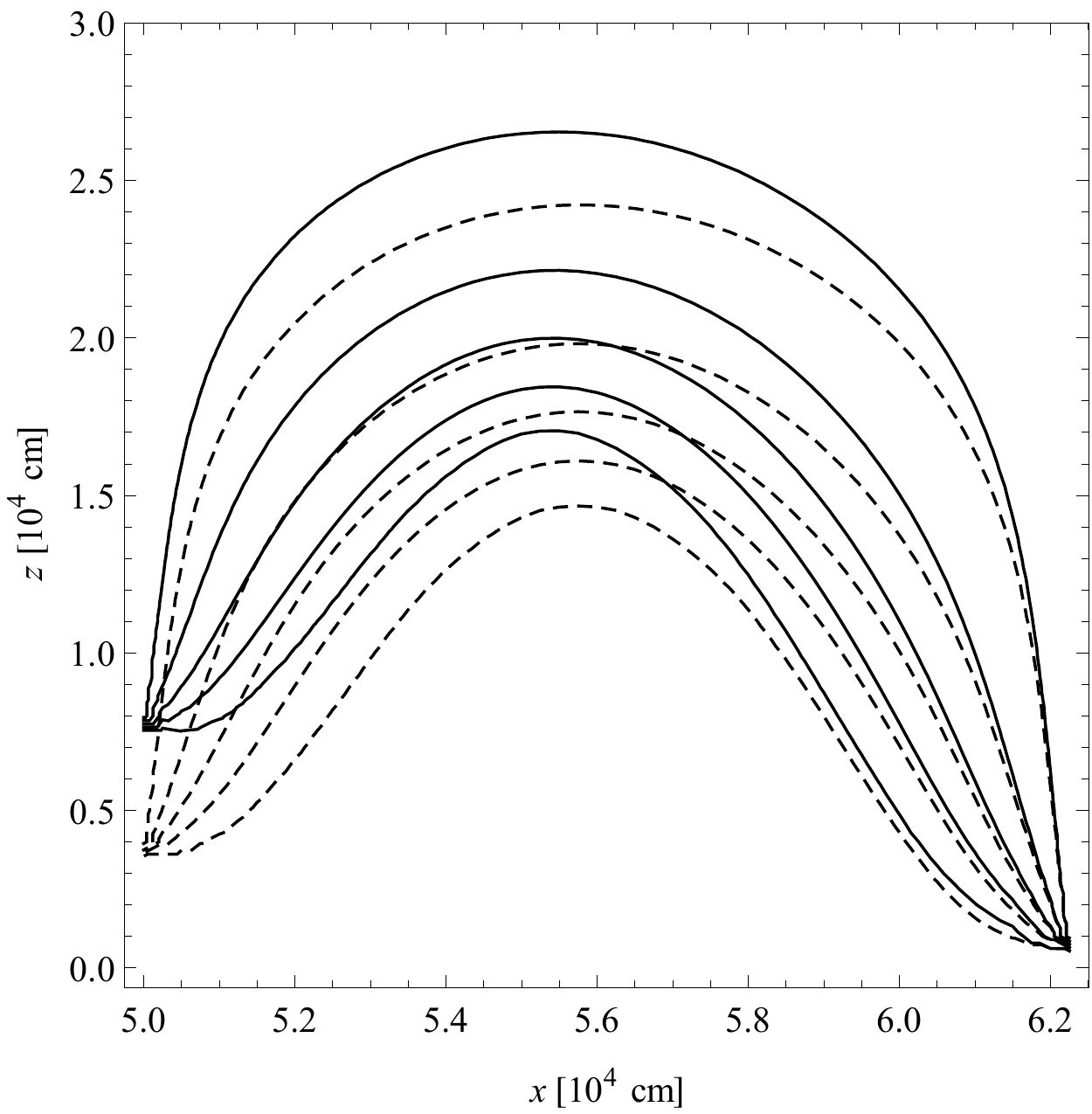}\\(b)
	\caption{Two-dimensional slices of the 3D structure of the shocks in the $xz$ plane: the filled column (a)
and unclosed hollow column (b).  The contours of $Q$ are shown from 0.9 to 0.1 (from top to bottom) with interval
0.2 for $\alpha=15^\circ$ (dashed lines) and $\alpha=30^\circ$ (solid lines).
}
  \label{fig:slices}
	\end{center}
 \end{figure}

Two-dimensional slices of the column structure in the $xz$ plane are shown in  Fig. \ref{fig:slices}.
They demonstrate graphically the extent of asymmetry of solutions in the case of the filled column (Fig. \ref{fig:slices}a)
and hollow unclosed column (Fig. \ref{fig:slices}b).

\begin{figure}
  	\begin{center}
  \includegraphics[width=0.45\textwidth]{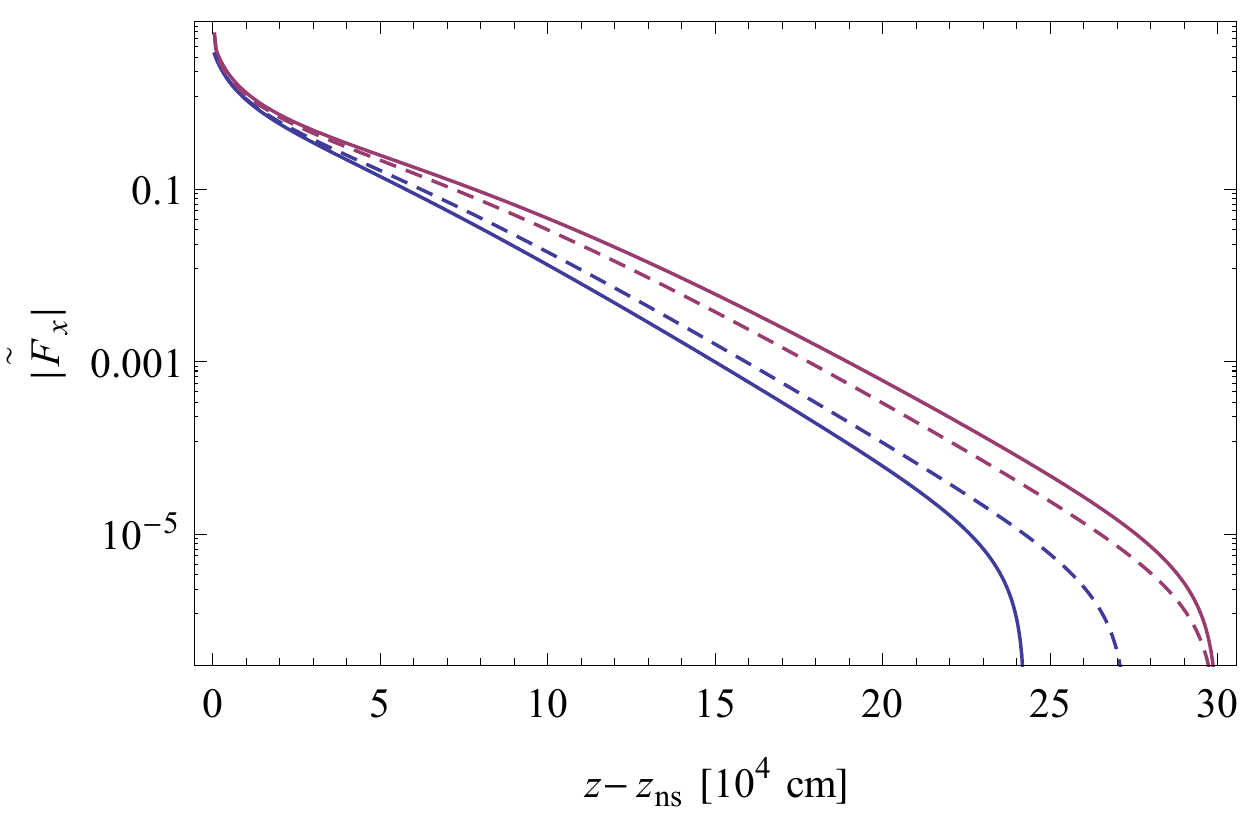}\\(a)\\\vspace{10pt}
  \includegraphics[width=0.45\textwidth]{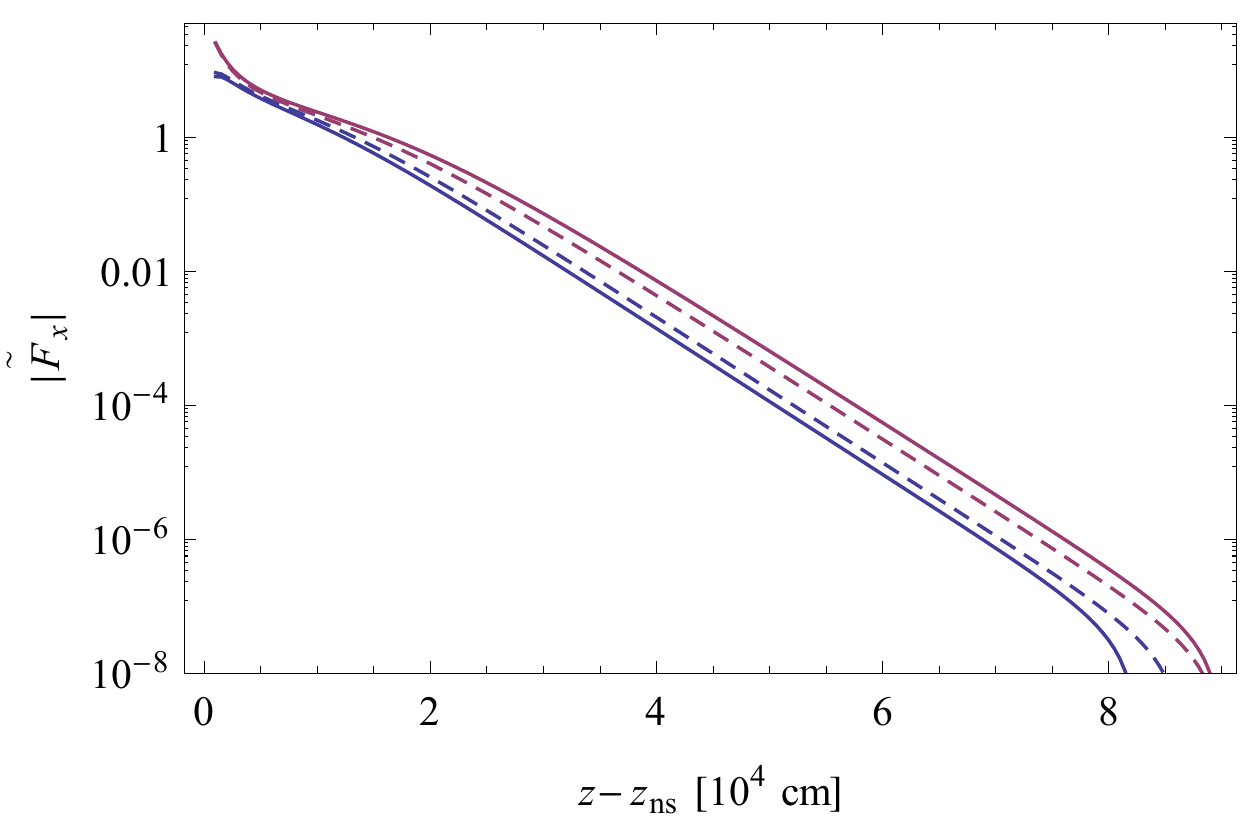}\\(b)
	\caption{The modulus of dimensionless component $\tilde{F}_x$ of radiative flux (\ref{e:flux})
plotted against the height $z-z_{\rm ns}$ above the neutron star surface in the $xz$ plane for the filled (a)
and unclosed hollow (b) column.  The dependencies are shown for $\alpha=15^\circ$ (dashed lines) and $\alpha=30^\circ$ (solid lines).
Two lower lines correspond to the left boundaries, and two upper ones,
to the right boundaries of 2D slices (see Fig. \ref{fig:slices}).}
  \label{fig:Fx}
	\end{center}
 \end{figure}

Fig. \ref{fig:Fx} shows calculated in the $xz$ plane distributions of the modulus of the dimensionless component of the radiation flux
  $\tilde{F}_x=F_x/(S\varv_0^2/2)={\partial Q}/{\partial \tilde{x}}$ along the column sidewall.
  In the figures, $z_{\rm ns}$ denotes the vertical coordinate of the neutron star
  surface at current $x$ and $y$.
At fixed $\dot M$ and $A$, the solution for $Q$ does not depend on $\varv_0$
when the first-type boundary condition (\ref{e:boundSW1}) is set at the side surface.
Therefore, it is useful to represent the radiation flux in units of $S\varv_0^2/2$. Since $D^\perp\propto\varv_0\sqrt{Q}$,
the realization of conditions of the type (\ref{e:boundSW}) at the side boundary
introduces the dependence of the solution on $\varv_0$.
The variation of $\varv_0$ in the common-sense ranges entails the slight change of the shock width near the column sidewall.

The quantities
\beq{e:f1}
f(\alpha)=\frac{\int |{F}_x (-r_0, 0, z, \alpha)|{\rm d}z}{\int {F}_x(r_0, 0, z,\alpha) {\rm d}z}
\eeq
for the filled column and
\beq{}
f'(\alpha)=\frac{\int |{F}_x (r_0, 0, z, \alpha)|{\rm d}z}{\int {F}_x(r_0+br_0, 0, z, \alpha) {\rm d}z}
\eeq
in the circlet-like geometry may be considered
as the measures of the azimuthal anisotropy of the column emission in the dependence on the angle value
 at fixed accretion rate and ceteris paribus.
 For the presented calculations, the particular values are $f(15^\circ)\simeq 0.81$, $f(30^\circ)\simeq 0.59$,
 $f'(15^\circ)\simeq 0.76$, and $f'(30^\circ)\simeq 0.60$.
In the case of the boundary condition (\ref{e:boundSW1}), when the quantity $Q$ at the sidewalls does not depend on the height,
the emergent flux  is calculated at the internal nodes of the grid closest to the boundary nodes (without
using the latter). This allows to avoid distortions near the base of the column: in the
immediate vicinity of the boundary, the radiation energy density behaves qualitatively
as if condition of type (\ref{e:boundSW}) were specified there.

It is possible to define similar relations through the integral emission of the column sidewalls.
Let  $\Gamma_1$ and $\Gamma_2$ be the sections of the filled column sidewall that are separated by the $yz$ plane and thus lie in
different half-spaces, $x<0$ and $x>0$, respectively.
Moreover, let $\Sigma_1$ denote the surface of the interior sidewall of the hollow column
(corresponding to the radius $r_0$) and $\Sigma_2$, the surface of the outer one.
Then, the ratio of the luminosities of the specified surfaces in each case will read as
\beq{e:ell1}
\ell(\alpha)=\frac{\iint_{\Gamma_1} {F}_n (x, y, z, \alpha){\rm d}\Gamma_1}{\iint_{\Gamma_2} {F}_n(x, y, z,\alpha) {\rm d}\Gamma_2}
\eeq
and
\beq{e:ell2}
\ell'(\alpha)=\frac{\iint_{\Sigma_1} {F}_n (x, y, z, \alpha){\rm d}\Sigma_1}{\iint_{\Sigma_2} {F}_n(x, y, z, \alpha) {\rm d}\Sigma_2}.
\eeq
In all expressions (\ref{e:f1})-(\ref{e:ell2}) the dependence of the flux component on the angle $\alpha$ should be understood as the
dependence on the problem parameter determining the geometry of the computational area and affecting the solution.

\begin{figure}
  	\begin{center}
  \includegraphics[width=0.45\textwidth]{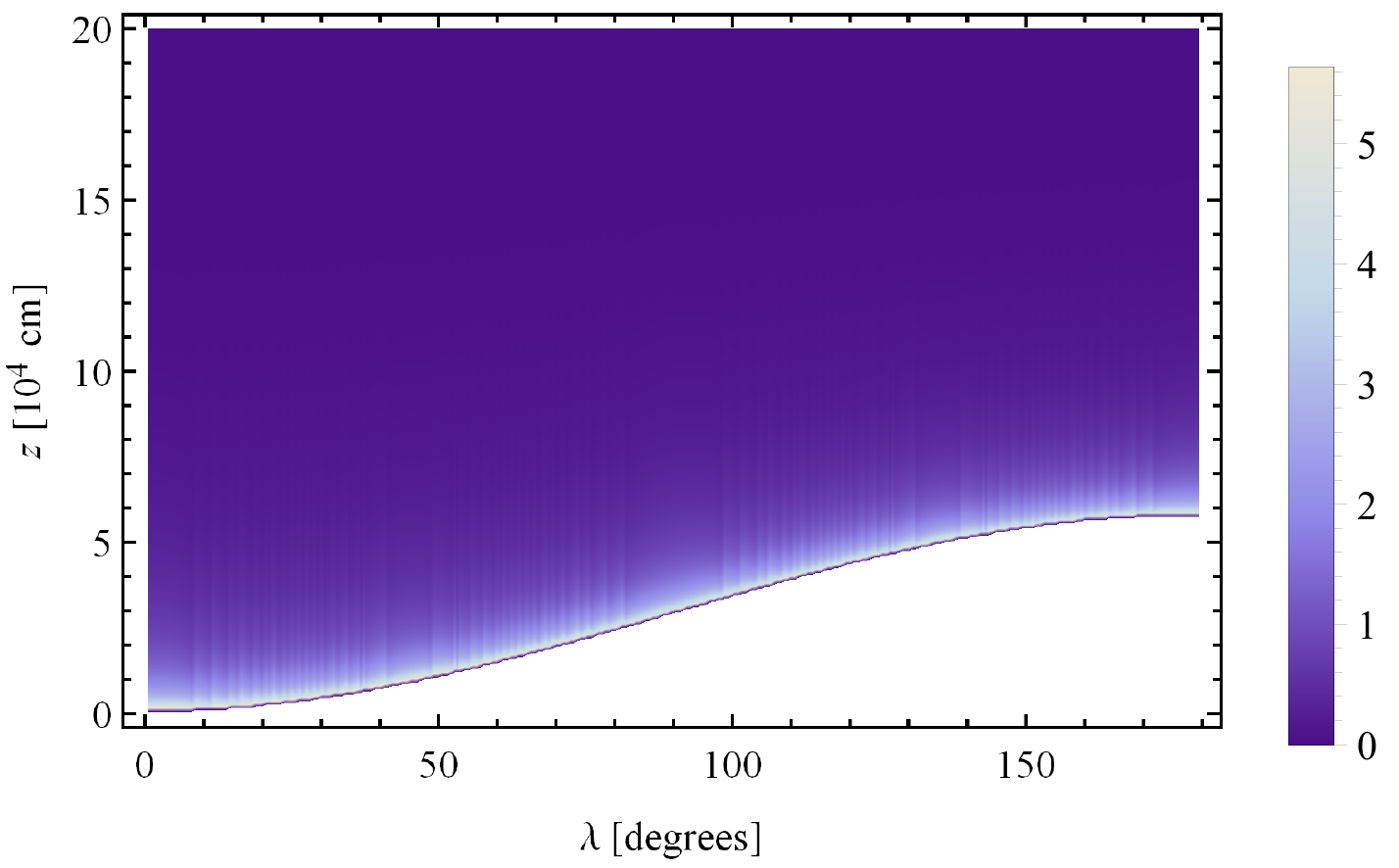}\\(a)\\\vspace{10pt}
  \includegraphics[width=0.45\textwidth]{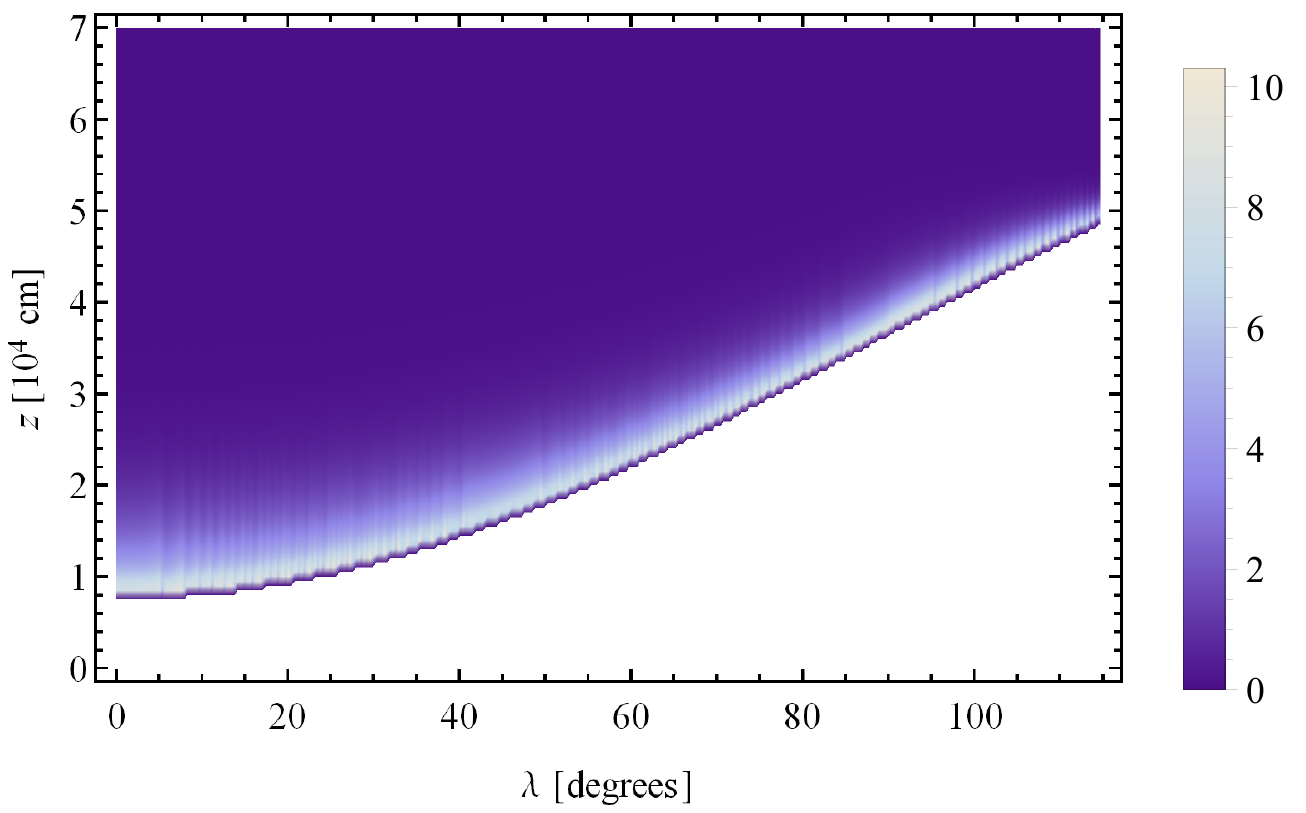}\\
  \includegraphics[width=0.45\textwidth]{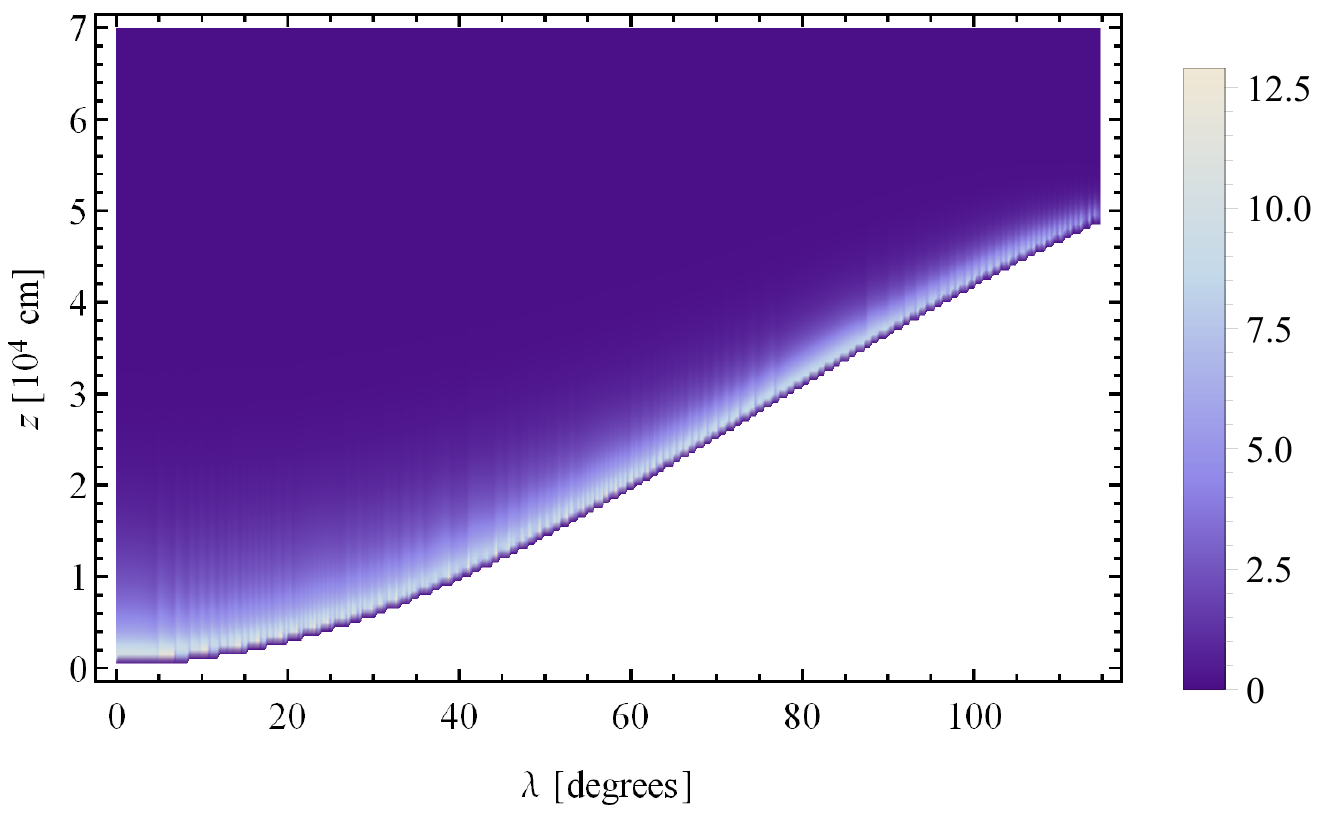}\\(b)
	\caption{The distributions of the normal component of radiation flux $\tilde{F}_n$ over the
column sidewalls  for the filled (a) and unclosed hollow (b) geometry, $\alpha=30^\circ$. The upper of the panels (b) corresponds
to the inner sidewall, and the bottom panel, to the outer one.}
  \label{fig:Fn}
	\end{center}
 \end{figure}

The calculations give $\ell(15^\circ)\simeq 0.92$, $\ell(30^\circ)\simeq 0.83$,
$\ell' (15^\circ)\simeq 0.95$ and $\ell'(30^\circ)\simeq 0.87$.
Fig.~\ref{fig:Fn} displays 2D distributions of $\tilde{F}_n$ (normalization is prior) over the column surface.
The angular coordinate
$\lambda$ is the azimuthal angle counted in the $xy$ plane.
The panels in Fig.~\ref{fig:Fn}b  do not show the regions with $\lambda>115^\circ$ where the channel
is too thin to linger the radiation in the medium long enough and stop the flow at a significant height:
here the plasma falls freely nearly to the neutron star surface causing almost zero diffusion flux from the side boundary.

The solutions illustrate the importance of taking into account the radiation diffusion across the magnetic field,
accompanied by the process of advection in the vertical direction.
The main contribution to the luminosity is caused by the photons coming from
the side surface of the column not very far from neutron star surface and forming a fan beam.
The investigation of the properties of observed pulse profiles is a problem deserving a separate consideration.
To compare theoretical results with observational data and construct the conclusions concerning the influence of spatial
asymmetry, one should take into account the angular distribution of the sidewall emission which is directed mainly
towards the neutron star surface.
The questions of the emergent radiation spectra and the distribution of spectral radiation flux over the column surface
will be considered in the grey approximation in the following section.

\section{The photon energy as a third coordinate}
\label{sec:3dsp}
\subsection{The equations and boundary conditions}
The spatial and time evolution of the angle-averaged photon occupation number  in the moving compressible medium
can be described in the diffusion approximation by the kinetic equation  (\citealt{1981MNRAS.194.1033B}, \citealt{1982SvAL....8..330L})
\beqa{e:BP}
\nabla\cdot\left(D \nabla n\right)- \bm \varv\cdot\nabla n
+ \frac{n_{\rm e}\sigma_{\rm T}}{m_{\rm e}c \epsilon^2}\frac{\partial}{\partial\epsilon}
\left[\epsilon^4\left(n+kT_{\rm e}\frac{\partial n}{\partial \epsilon}\right)\right]\nonumber\\
+ \nabla\cdot\bm \varv\frac{\epsilon}{3}\frac{\partial n}{\partial\epsilon}=\frac{\partial n}{\partial t},
\eeqa
from which equation  (\ref{e:kinint}) and, as a consequence, equation (\ref{e:Qeq}) are derived. Equation (\ref{e:BP}) takes into account the diffusion, advection, and thermal and bulk Comptonization.
To describe the thermal Comptonization in strongly magnetized plasma, the mean cross-section $\langle \sigma \rangle$ detemined by the cross-sections along and across the magnetic field can be introduced \citep{2007ApJ...654..435B}.

Now consider the axially symmetric filled column in the model of circle cylinder.
Introducing  cylindrical coordinates centred at the axis of the cylinder so that
the $z$ axis is again directed towards the flow and  $z=0$ at the neutron star surface, let us now solve in these coordinates the system
including  \eq{e:BP}, that turns out to be 3D in the $rz\epsilon$ space, instead of equation (\ref{e:energy1}).
Thus the problem is to solve the stationary system
of equations  (\ref{e:BP}), (\ref{e:momentum}), (\ref{e:cont}), (\ref{e:Te}) (${\partial n}/{\partial t}=0$),
which is reduced to the system  (\ref{e:BP}), (\ref{e:r-density}), (\ref{e:Te}) when the gravitation is
neglected (the present case).

The velocity (\ref{e:r-density}) is determined  making use of the quantity  $u=\int u_\epsilon d\epsilon$, where
$u_\epsilon=8\upi\epsilon^3n/(c^3h^3)$. The expression for the temperature (\ref{e:Te})
is a consequence of solving the Fokker--Planck equation for thermal electrons having a stationary distribution function and
 interacting with non-equilibrium radiation which is assumed to be isotropic.

At the upper boundary, at the height $z=z_0$, the velocity component $\varv_z=-\varv$
is determined by the free-fall velocity value,
\beq{e:momboundary}
\varv_z(r, z_0)=-\varv_0.
\eeq

The components of the spectral radiation flux are written
as follows:
\beq{e:Fr}
F_{\epsilon,\,r}=-D^\perp \frac{\partial u_{\epsilon}}{\partial r},
\eeq
\beq{e:Fz}
F_{\epsilon,\,z}=-D^\| \frac{\partial u_{\epsilon}}{\partial z}+\varv_z u_{\epsilon}
- \frac{\varv_z\epsilon}{3}\frac{\partial u_{\epsilon}}{\partial\epsilon}.
\eeq
Free  escape of the photons from the column leads again to the following condition at the sidewall:
\beq{e:rightbound}
F_{\epsilon,\,r}(r_0, z)=\frac{2}{3}cu_{\epsilon}(r_0, z).
\eeq
Since the solution of the momentum equation is written under the condition $u(r, z_0)=0$, at the upper boundary it is reasonable to set
\beq{e:upbound}
u_{\epsilon}(r, z_0)=0.
\eeq
The condition
$-D^\| \frac{\partial u_{\epsilon}}{\partial z}=\frac{2}{3}cu_{\epsilon}$
is also appropriate and becomes 
equivalent to condition (\ref{e:upbound}) far above the shock, where the velocity
is approximately constant.
At the central axis
\beq{e:leftbound}
\frac{\partial u_{\epsilon} (0,z)}{\partial r}=0.
\eeq

The physical meaning of these relations corresponds to their frequency-integrated analogues.
At the bottom boundary, the approximate condition  $u(r, 0)=3S\varv_0$ is used, which means that $\varv(r, 0)=0$.
Then at this boundary one can set
\beq{e:bbtemperature}
a T_0^4 = u(r, 0),
\eeq
where $a=8\upi^5k^4/(15h^3c^3)$
is the radiation constant.
Determining from here $T_0$ one can suppose the blackbody bottom boundary condition for spectral radiation energy density:
\beq{e:bottomcondition}
u_{\epsilon}(r, 0)=\frac{8\upi\epsilon^3}{c^3h^3}\frac{1}{\exp\left(\frac{\epsilon}{kT_0}\right)-1}.
\eeq

It is also  assumed  \citep{2012A&A...538A..67F} that there are no photons at the boundaries of the photon energy grid,
$\epsilon_1=0.1~{\rm keV}$ and $\epsilon_2=500~{\rm keV}$,
so that
\beq{}
u_\epsilon(\epsilon_1)=u_\epsilon(\epsilon_2)=0.
\eeq

\subsection{Computations}

The stationary solution of the system described above is constructed by the iterative procedure based on the time relaxation method.
At each time-step, the electron temperature, velocity, and spectral radiation energy density are determined alternatively by expressions
(\ref{e:Te}), (\ref{e:r-density}),  and from \eq{e:BP}, respectively. As the initial distribution of the spectral radiation energy
density, the blackbody spectrum with temperature $0.1$ keV is used.

To realize the numerical modelling, it is reasonable to introduce the new variables.
Using the logarithmic scale, let us determine the photon energy as the function of the dimensionless
variable $\xi$, so that  $\epsilon=\epsilon_0 {\rm e}^\xi$, where $\epsilon_0=1~{\rm keV}$.
Then, the spectral radiation energy density per unit of $\xi$ is equal to
\beq{}
u_{\xi}=u_{\epsilon} \frac{{\rm d}\epsilon}{{\rm d}\xi}=\frac{8\upi\epsilon_0^4{\rm e}^{4\xi}}{c^3h^3}n.
\eeq
For computational reasons, it is convenient to write the equations for the function $\tilde{u}_{\xi}={\rm e}^{4\xi}n$.
The dimensionless variables are now defined by the relations:
\beqa{}
\tilde{r}=\frac{r}{R_0},~\tilde{z}=\frac{z}{R_0},~R_0=10^4~{\rm cm};\\\nonumber
\tilde{\varv}_z=\frac{\varv_z}{\varv_0};~
\theta=\frac{kT_{\rm e}}{\epsilon_0};~
\tilde{t}=\frac{t}{t_0},~t_0=\frac{\sigma_{\rm T} \dot{M}R_0^2}{ \varv_0 c m_{\rm p} \upi r_0^2 }.
\eeqa
In terms of these variables \eq{e:BP} can be written as follows:
\beqa{e:BPdim}
\frac{1}{\tilde{r}}\frac{\partial}{\partial\tilde{r}}
\left(\tilde{r}\tilde{D}^\perp\frac{\partial\tilde{u}_{\xi}}{\partial\tilde{r}} \right)
+\frac{\partial}{\partial\tilde{z}}
\left(\tilde{D}^\|\frac{\partial\tilde{u}_{\xi}}{\partial\tilde{z}} \right)
-\frac{\varv_0t_0}{R_0}\tilde{\varv}_z\frac{\partial\tilde{u}_{\xi}}{\partial \tilde{z}}\nonumber\\
+\frac{t_0 f_0\langle\tilde{\sigma}\rangle}{|\tilde{\varv}_z|}\theta\frac{\partial^2 \tilde{u}_{\xi}}{\partial\xi^2}
+\left(\frac{1}{3}\frac{\varv_0t_0}{R_0}\frac{\partial\tilde{\varv}_z}{\partial\tilde{z}}
+\frac{t_0 f_0\langle\tilde{\sigma}\rangle}{|\tilde{\varv}_z|}
\left({{\rm e}^{\xi}}-5\theta\right)\right)\frac{\partial\tilde{u}_{\xi}}{\partial\xi}\nonumber\\
+\left(\frac{t_0 f_0\langle\tilde{\sigma}\rangle}{|\tilde{\varv}_z|}4\theta
-\frac{4}{3}\frac{\varv_0t_0}{R_0}\frac{\partial\tilde{\varv}_z}{\partial\tilde{z}} \right) \tilde{u}_{\xi}
=\frac{\partial \tilde{u}_{\xi}}{\partial \tilde{t}}.
\eeqa
Here,
\beqa{}
\tilde{D}^{\perp, \|}=\frac{t_0}{R_0^2}D^{\perp, \|},
~\langle\tilde{\sigma}\rangle=\frac{\langle\sigma\rangle}{\sigma_{\rm T}}=\sqrt{\frac{D^{\perp}}{ D^{\|}}},\\\nonumber
\phi_0=\frac{c\sigma_{\rm T}\dot{M}}{\varv_0m_{\rm p}\upi r_0^2},~
f_0=\frac{\epsilon_0}{m_{\rm e}c^2}\phi_0,\nonumber
\eeqa
the quantities $D^{\perp}$ and $D^{\|}$ are determined by expressions (\ref{e:Dperp}) and (\ref{e:Dpar}), respectively
(the ratio $D^{\|}/D^{\perp}$ will be varied).
The quantity $\langle \sigma \rangle$ in calculations is supposed to be the geometrical mean of the  values of cross-sections in the directions along and across the magnetic field.

The main computational problem is the solution (on each time step) of the \eq{e:BP} (\eq{e:BPdim}).
Since the implicit alternating direction scheme (which was used, for example, in the work of \cite{2012A&A...538A..67F} for the case `1 spatial coordinate + photon energy') in 3D case (`2 spatial coordinates + photon energy') is generally speaking
unstable, it is possible to apply the locally 1D implicit scheme \citep{2001samarskii_eng}, which is based on the so-called method of summary approximation. However, this scheme  is only conditionally stable in the case of
considering equation (that can be shown, for example, with von Neumann method).
The explicit schemes implying parallel computations can also be used. During the creation of this work in different time,
both variants were being tested and used, depending on the exploited computational powers.
The results presented here are achieved making use of the rectangular, equidistant in dimensionless variables meshes and
the explicit scheme which includes the set of finite-difference operators
corresponding to the original equation (\ref{e:BPdim}).

The approximation of the
spatial diffusion  terms can be performed making use of the distinct finite-difference pattern functionals for the
calculation of the half-mesh value of the diffusion coefficient \citep{1962samarskii}.
Here, the arithmetical mean
is used (so that $D^{\perp,\|}_{i\pm 1/2}=(D^{\perp,\|}_i+D^{\perp,\|}_{i\pm 1})/2$, $i$ is the node number).
For example, the operator of spatial diffusion along the magnetic field is expressed making use of the finite-difference pattern
$\textsf{h}^{-2}\left(D^{\|}_{i+1/2}(\tilde{u}_{\xi, ~i+1}-\tilde{u}_{\xi, ~i})-D^{\|}_{i-1/2}(\tilde{u}_{\xi, ~i}-\tilde{u}_{\xi, ~i-1})\right)$, where $\textsf{h}$ is the mesh step.

The approximation of the diffusion terms is also possible after preliminary differentiation because of their continuity.
In this variant, second derivatives are approximated by central tree-point patterns, and first
derivatives in the terms
$\left(\frac{\tilde{D}^\perp}{\tilde{r}}
+\frac{\partial\tilde{D}^\perp}{\partial\tilde{r}}\right)\frac{\partial\tilde{u}_{\xi}}{\partial\tilde{r}}$ and
$\frac{\partial\tilde{D}^\|}{\partial\tilde{z}}\frac{\partial\tilde{u}_{\xi}}{\partial\tilde{z}}$ are approximated
by second-order central differences.

The second $\xi$ derivative is
approximated by a central tree-point pattern.
The terms containing the first $\xi$ derivatives are approximated making use of upwind differencing and accounting for general rules
leading to the conditional stability of the scheme.
The absence  of computational instabilities is controlled during the converging over the entire computational domain.

\subsection{Results}

 \begin{figure*}
 \hspace{20pt} \includegraphics[width=0.35\textwidth]{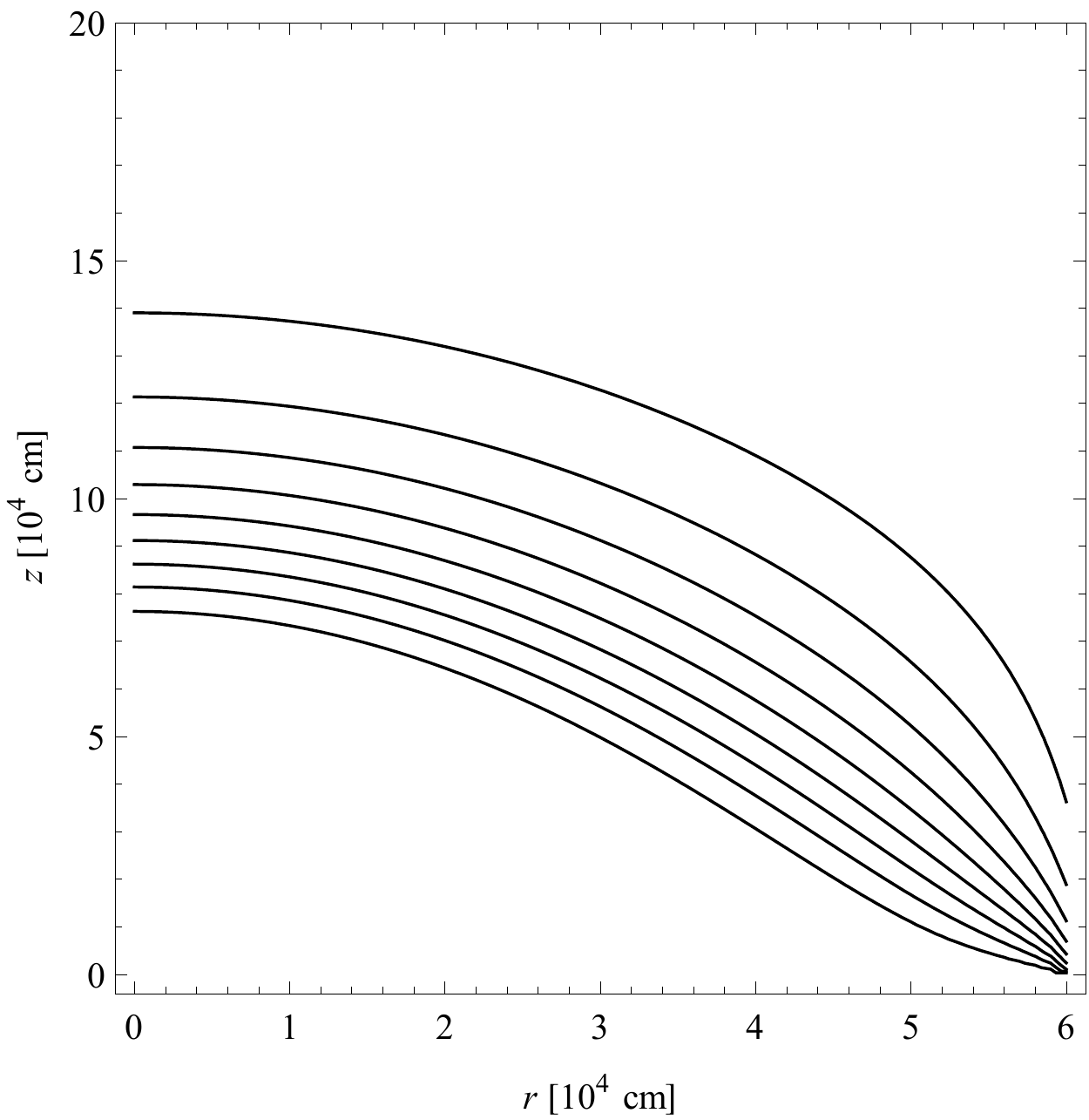}
  \hfill
                                \includegraphics[width=0.5\textwidth]{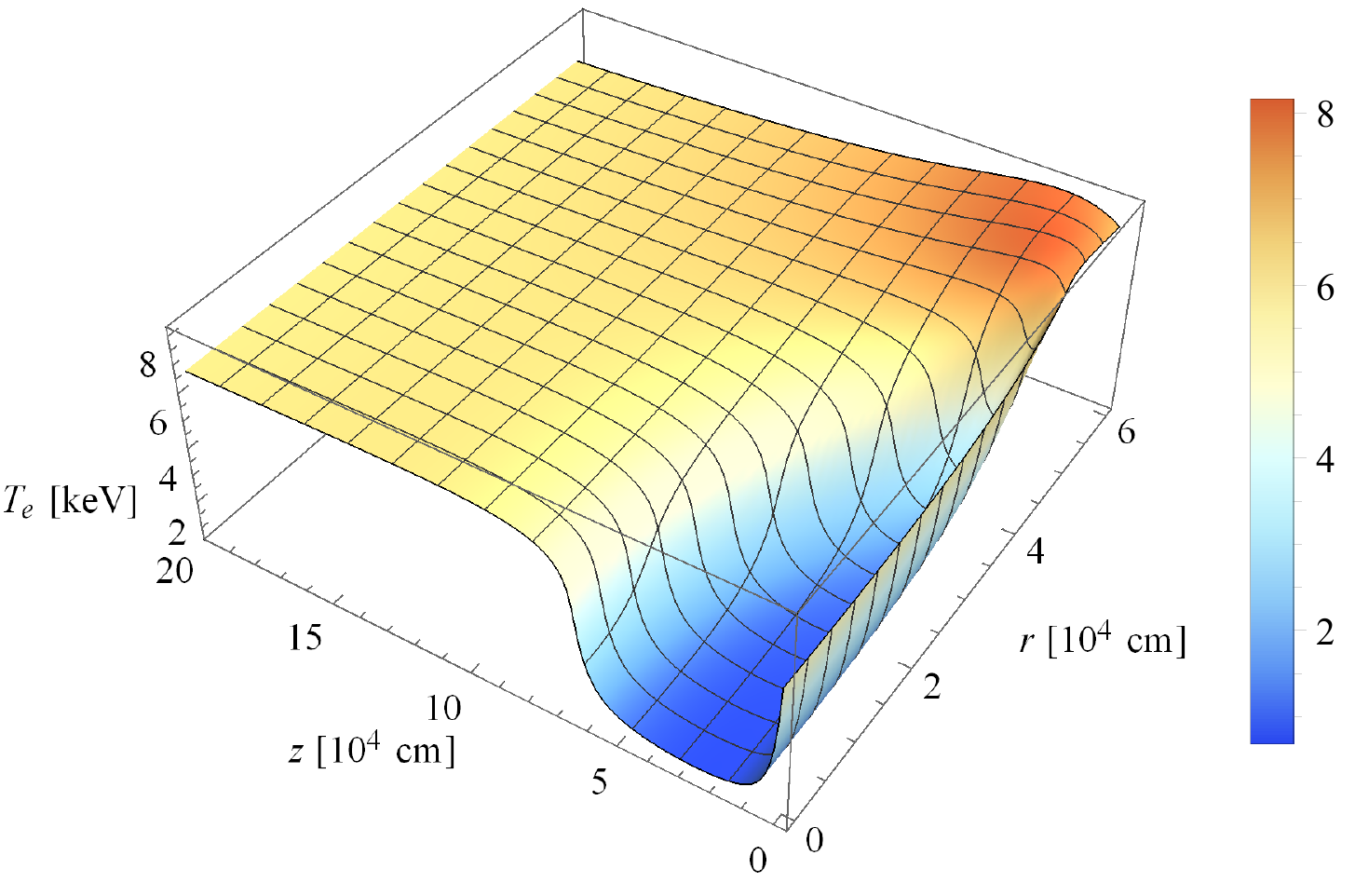}\vspace{10pt}
   \\
   (a) \hspace{270pt} (d)
  \\
  \vspace{10pt}          \hspace{20pt}  \includegraphics[width=0.35\textwidth]{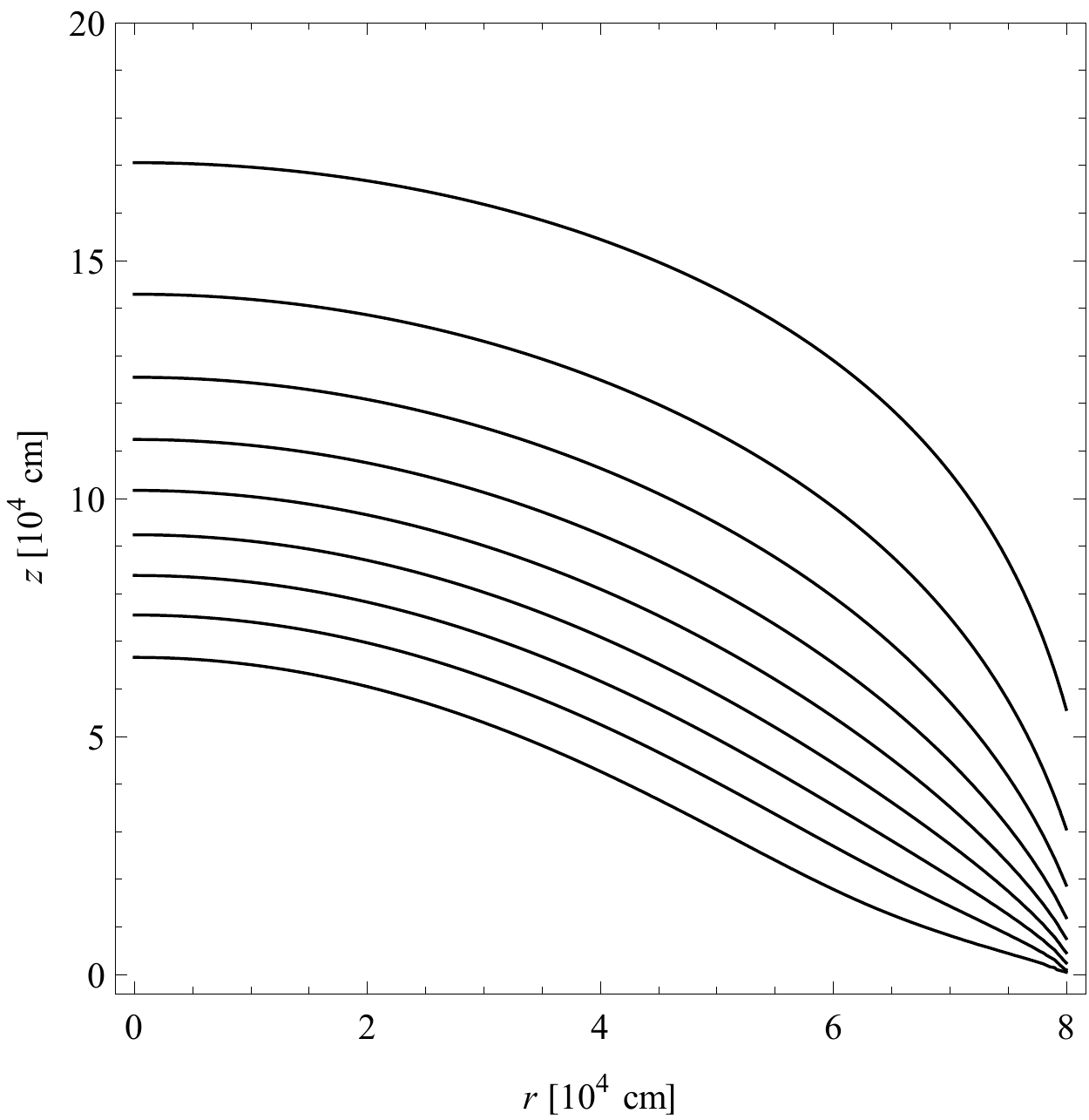}
\hfill
                                \includegraphics[width=0.5\textwidth]{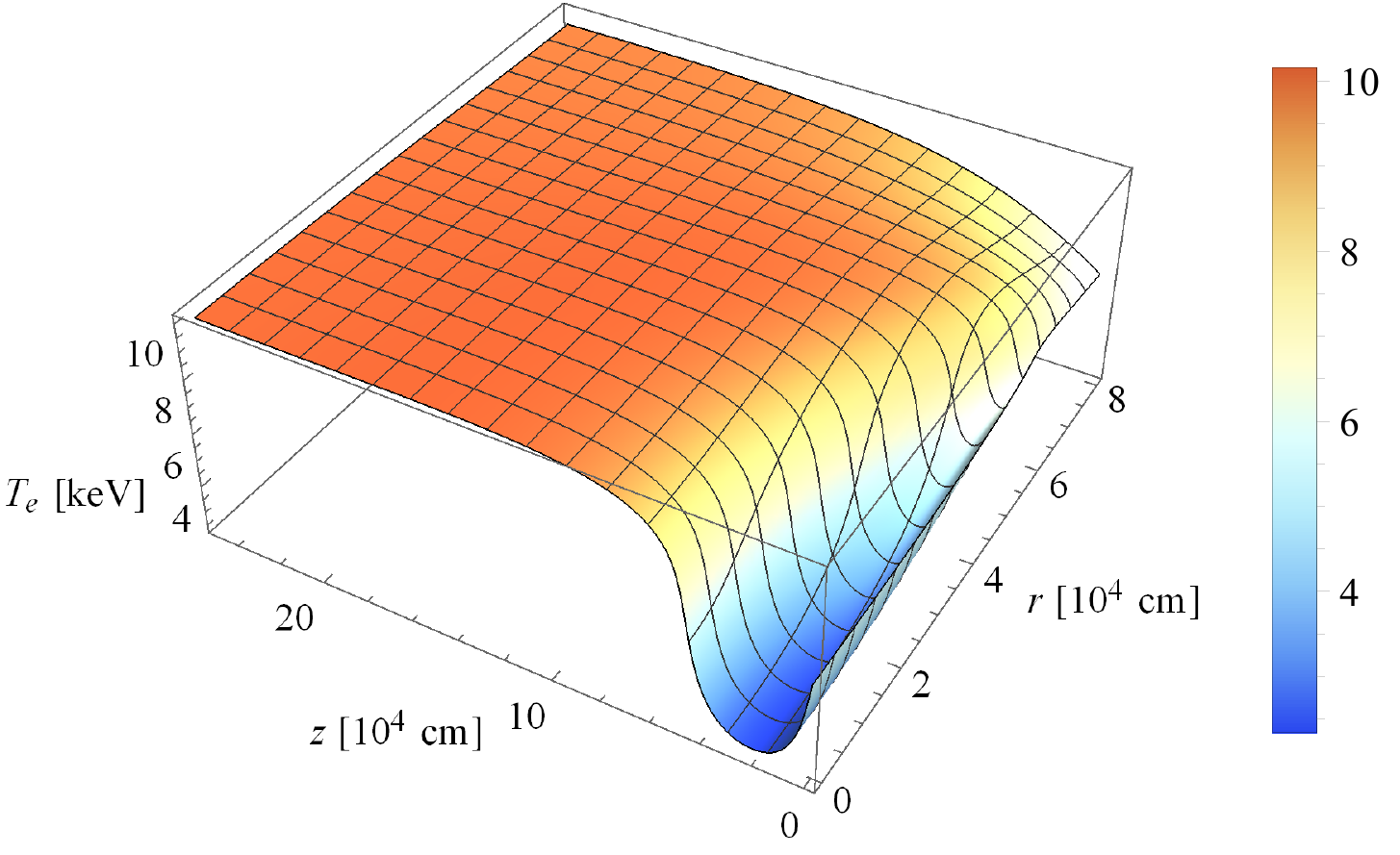}\vspace{10pt}
   \\
   (b) \hspace{270pt} (e)
  \\
  \vspace{10pt}

 \hspace{20pt}  \includegraphics[width=0.35\textwidth]{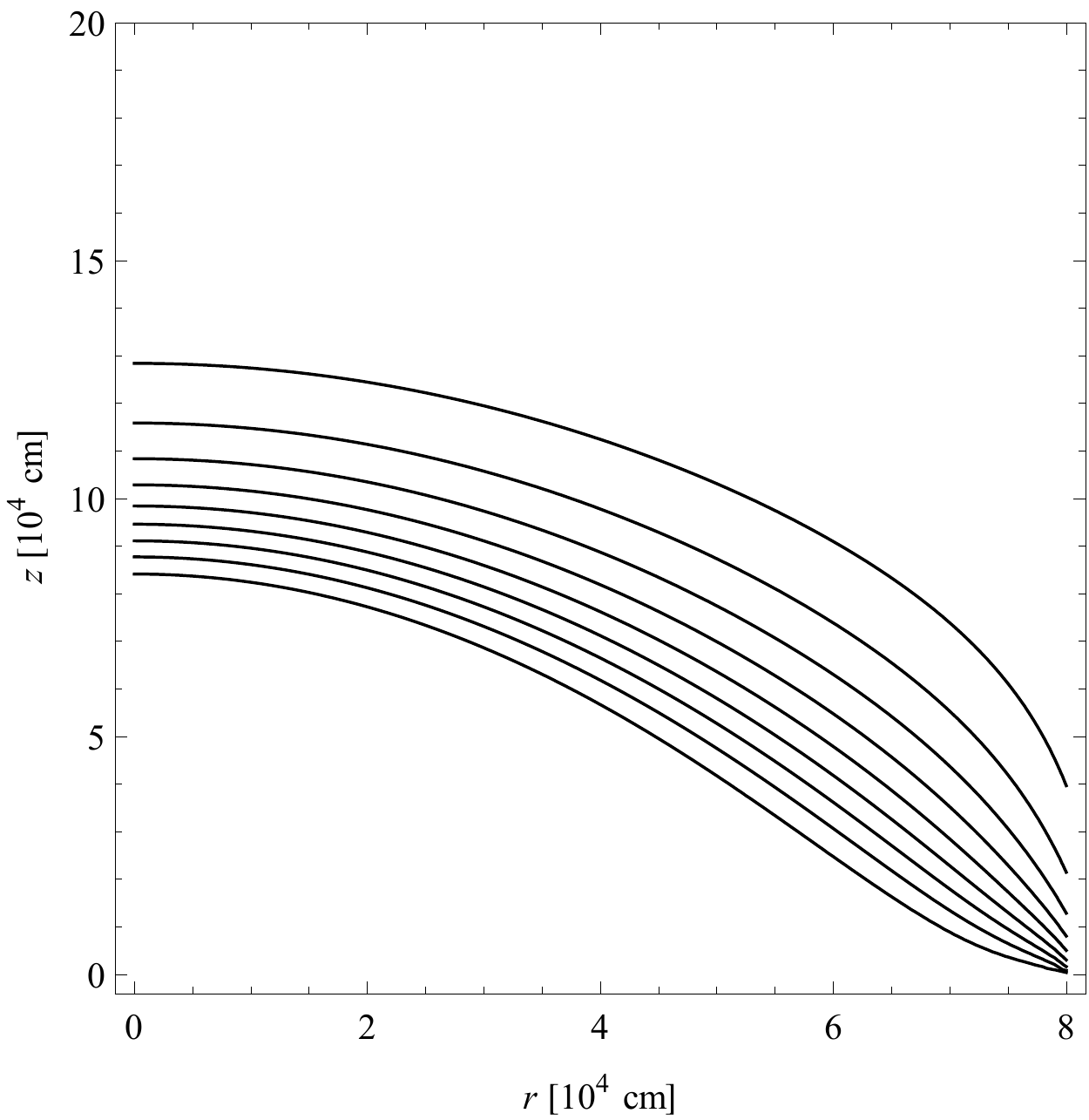}
  \hfill
                                \includegraphics[width=0.5\textwidth]{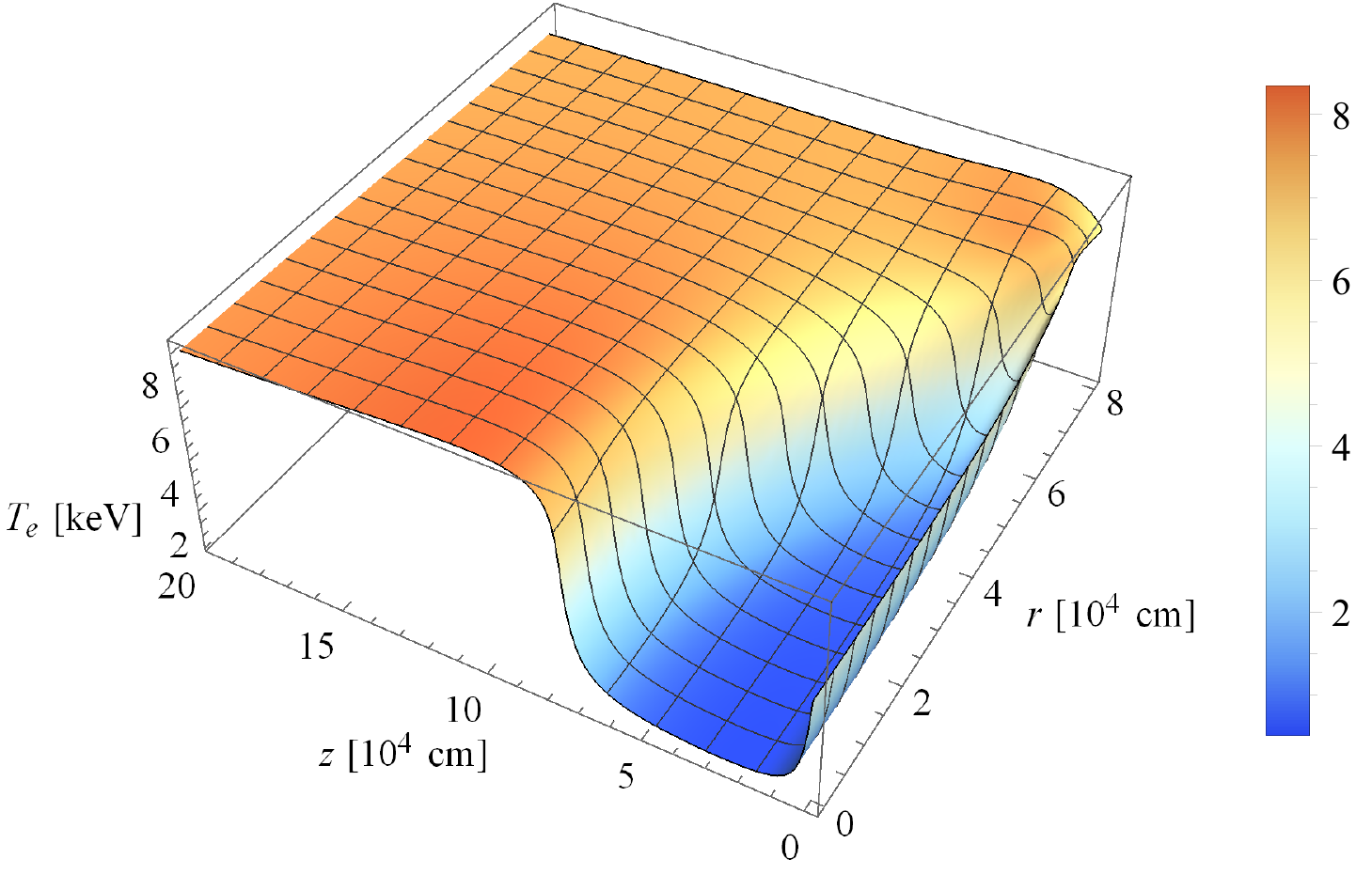}                                \\
 (c) \hspace{270pt} (f)
\caption{Contours of equal $Q$ (from 0.9 to 0.1 with interval 0.1, from top to bottom)
and profiles of the electron temperature (\ref{e:Te}) in the model of spatially 2D cylindrical accretion column,
that are the result of self-consistent modelling for different sets of model parameters: `1' (a, d),
`2' (b, e), `3' (c, f).}
\label{fig:spv}
 \end{figure*}

  \begin{figure*}
   \hspace{20pt}   \includegraphics[width=0.35\textwidth]{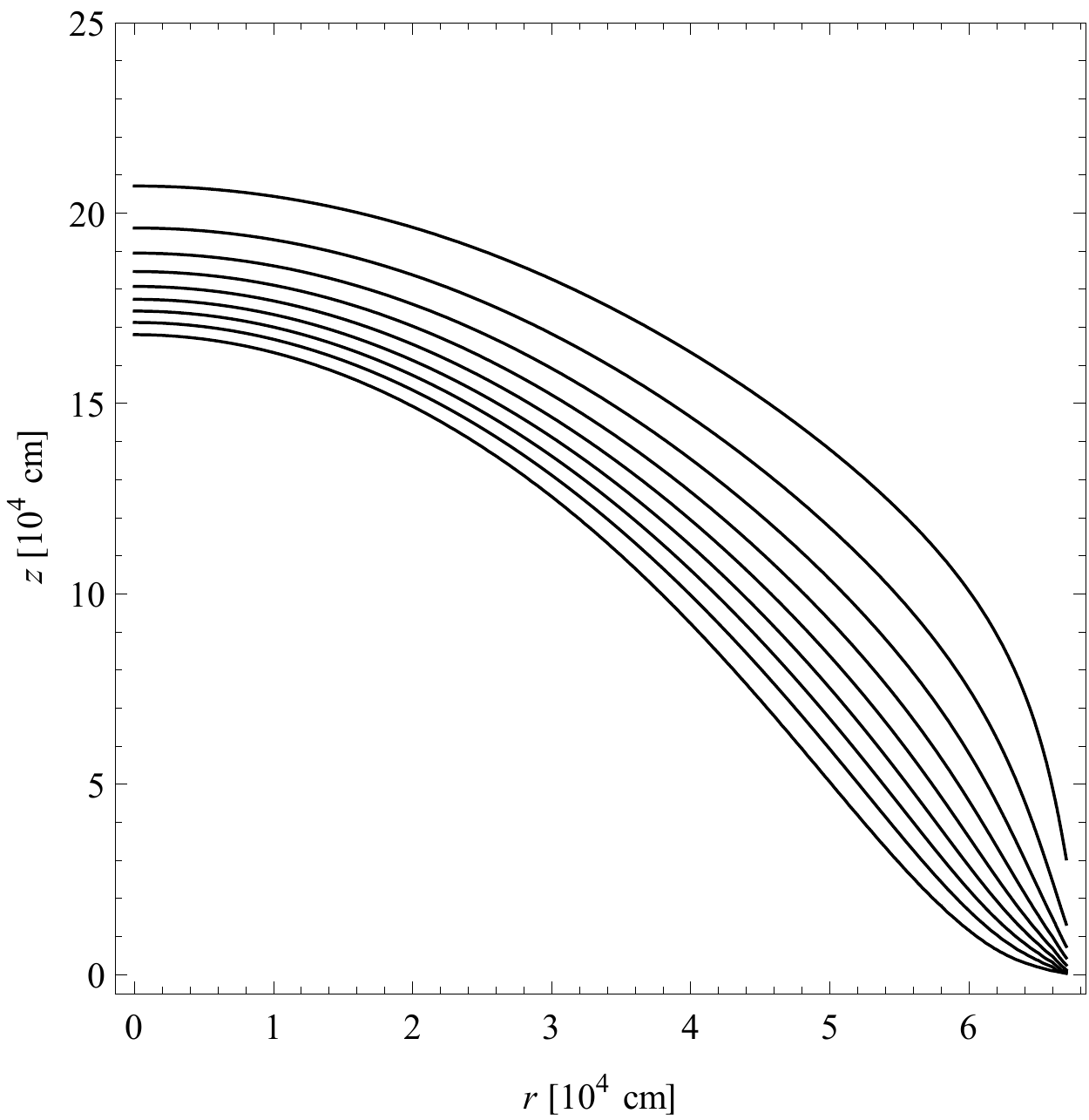}
  \hfill
                                 \includegraphics[width=0.5\textwidth]{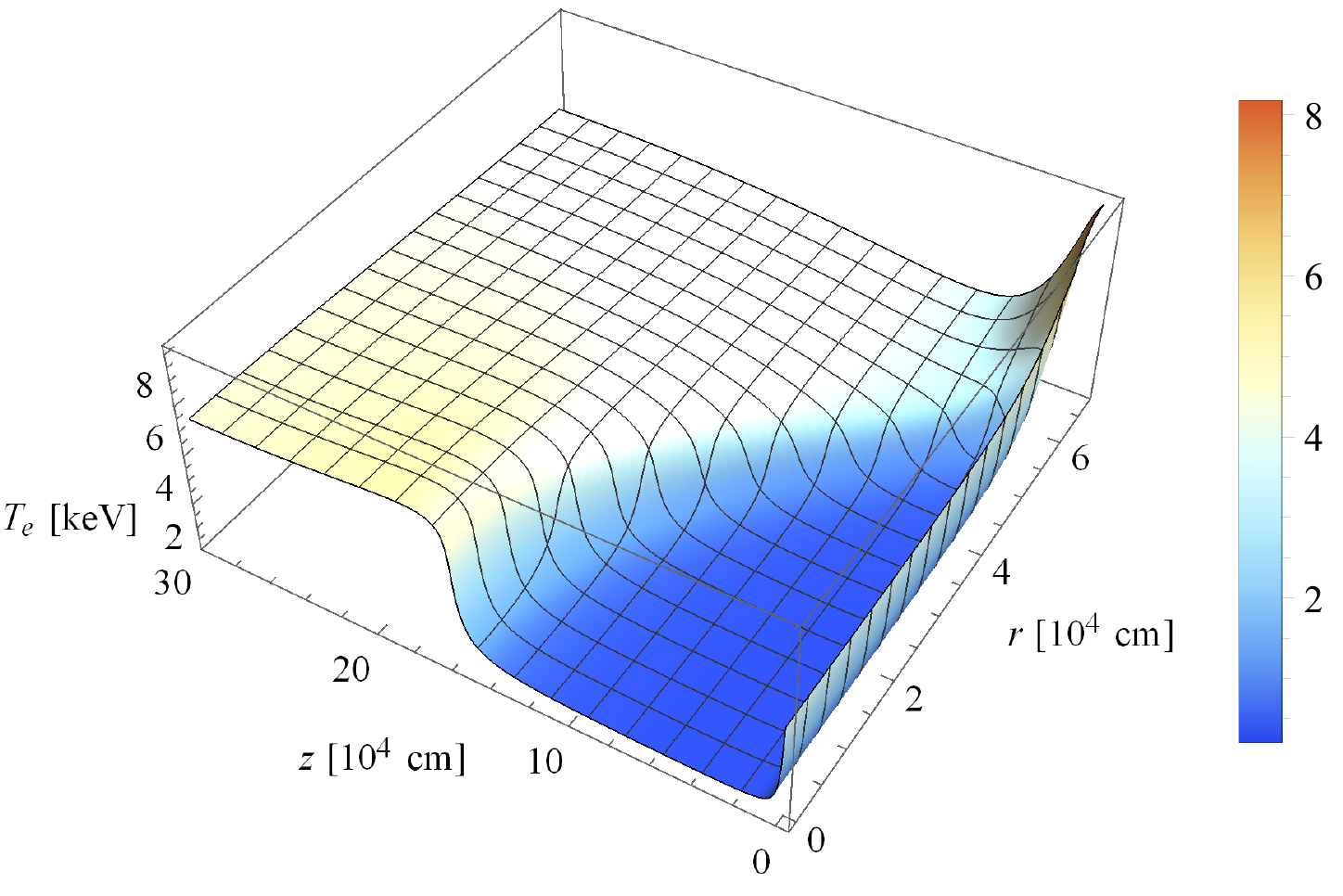}
   \\
   (a) \hspace{270pt} (d)
  \\
  \vspace{10pt}
  \hspace{20pt}   \includegraphics[width=0.35\textwidth]{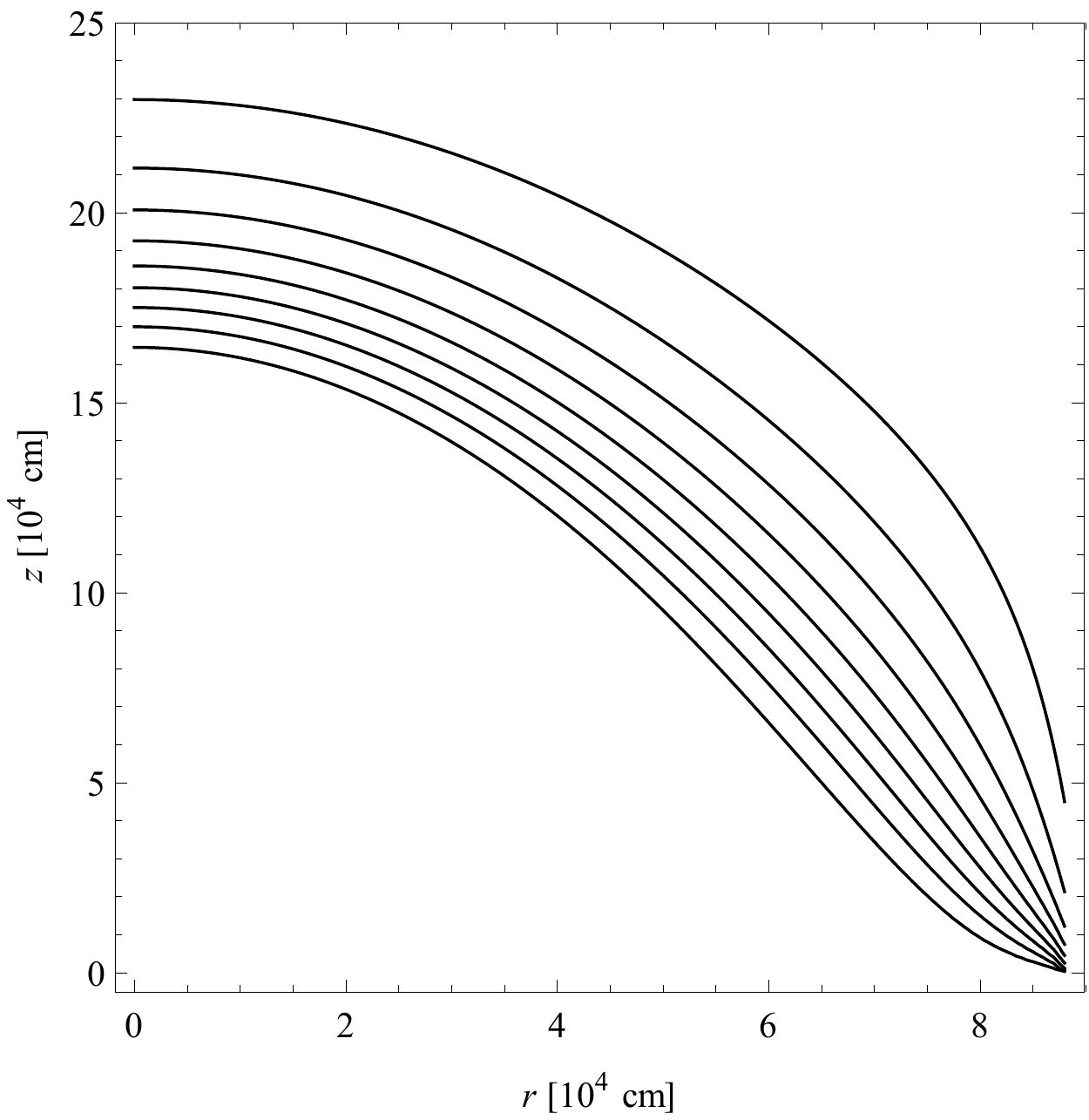}
  \hfill
                                \includegraphics[width=0.5\textwidth]{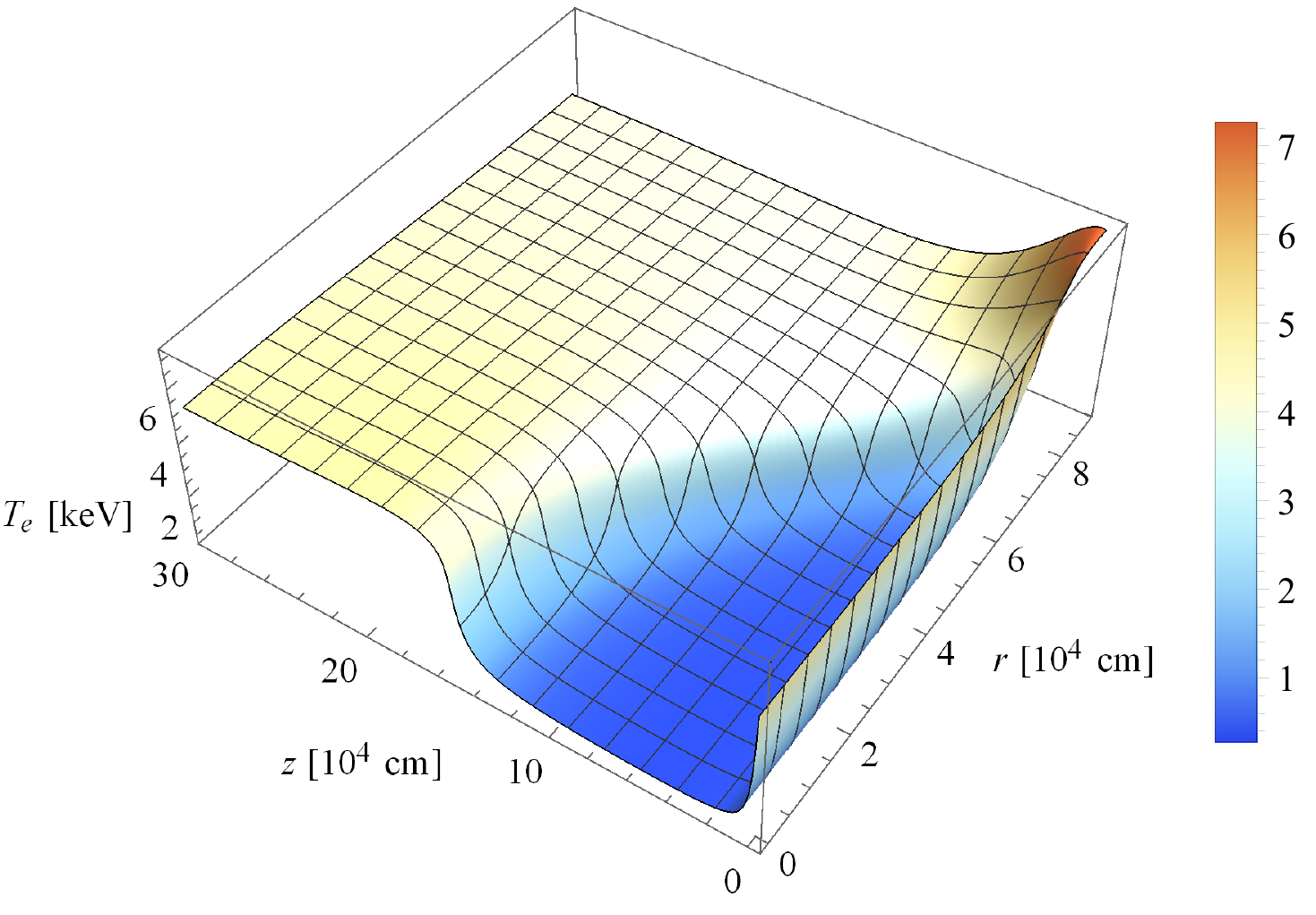}
     \\
     (b) \hspace{270pt} (e)
    \\
    \vspace{10pt}
 \hspace{20pt}  \includegraphics[width=0.35\textwidth]{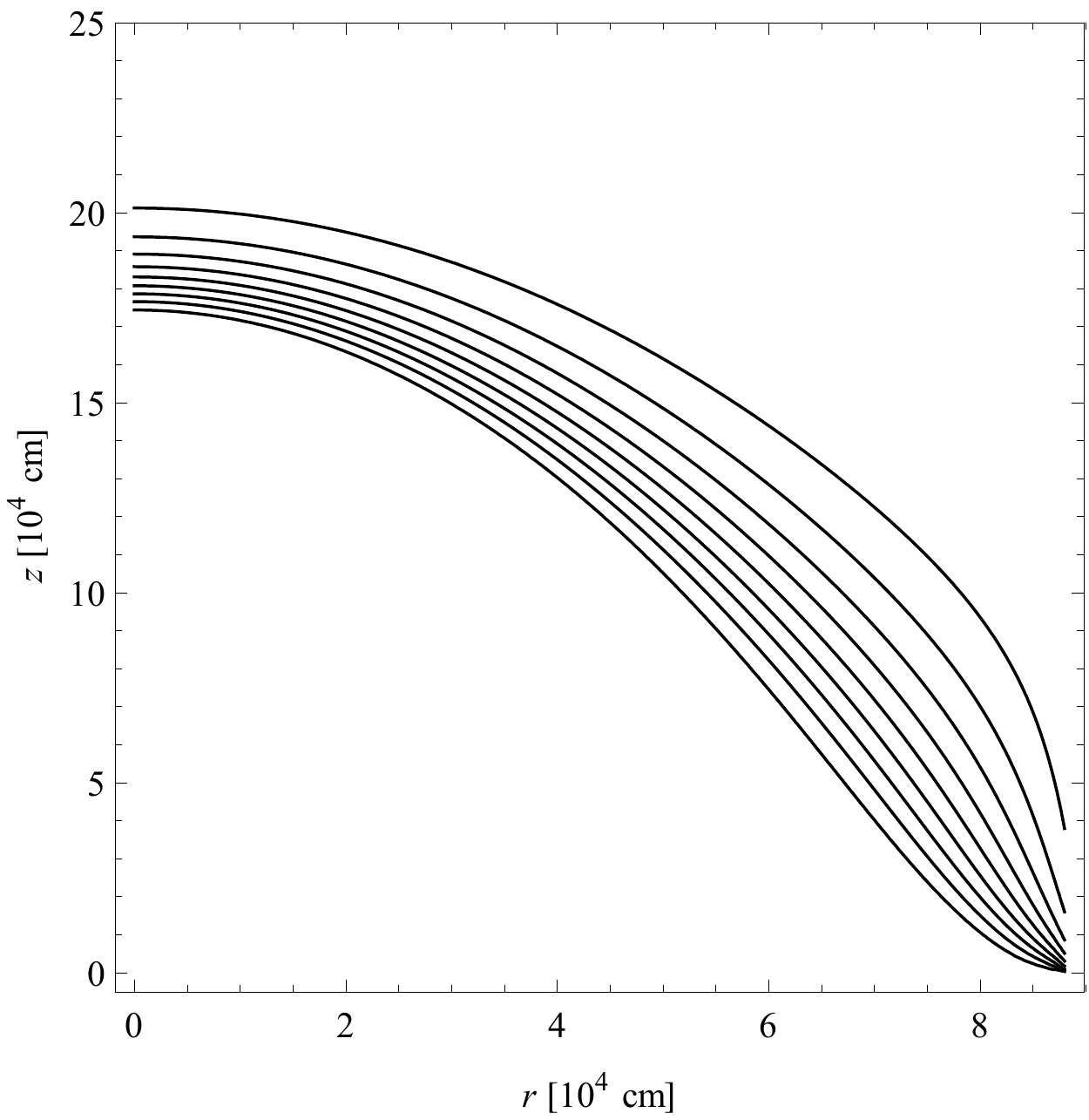}
  \hfill
    \includegraphics[width=0.5\textwidth]{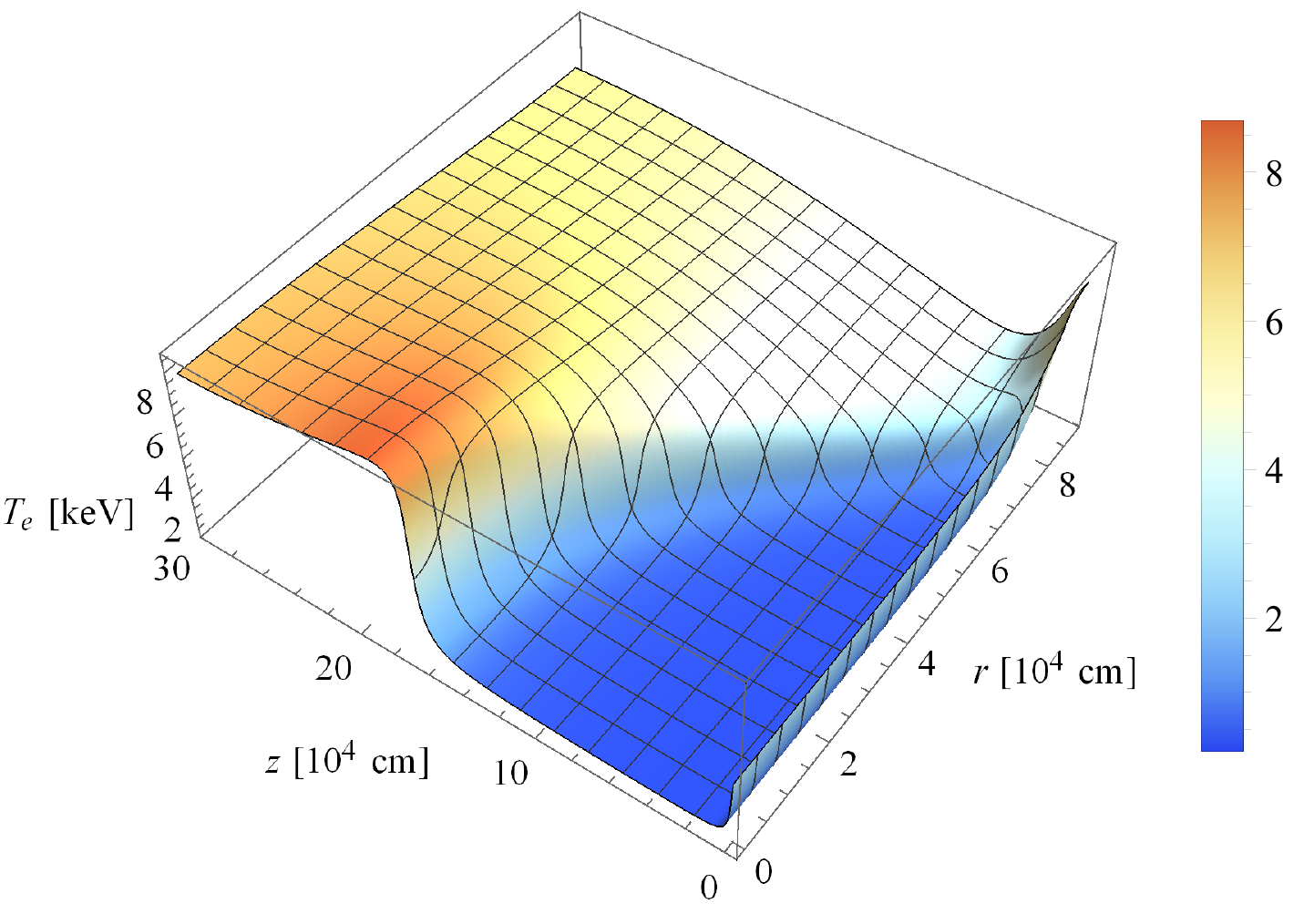}
  \\
  (c) \hspace{270pt} (f)

    \caption{Same as Fig. \ref{fig:spv}, but
for the mass accretion rate $\dot M_{17}=2$, for different sets of model parameters: `4' (a, d),
`5' (b, e), `6' (c, f).
}
  \label{fig:dotm2}
 \end{figure*}

\begin{figure}
  	\begin{center}
  \includegraphics[width=0.45\textwidth]{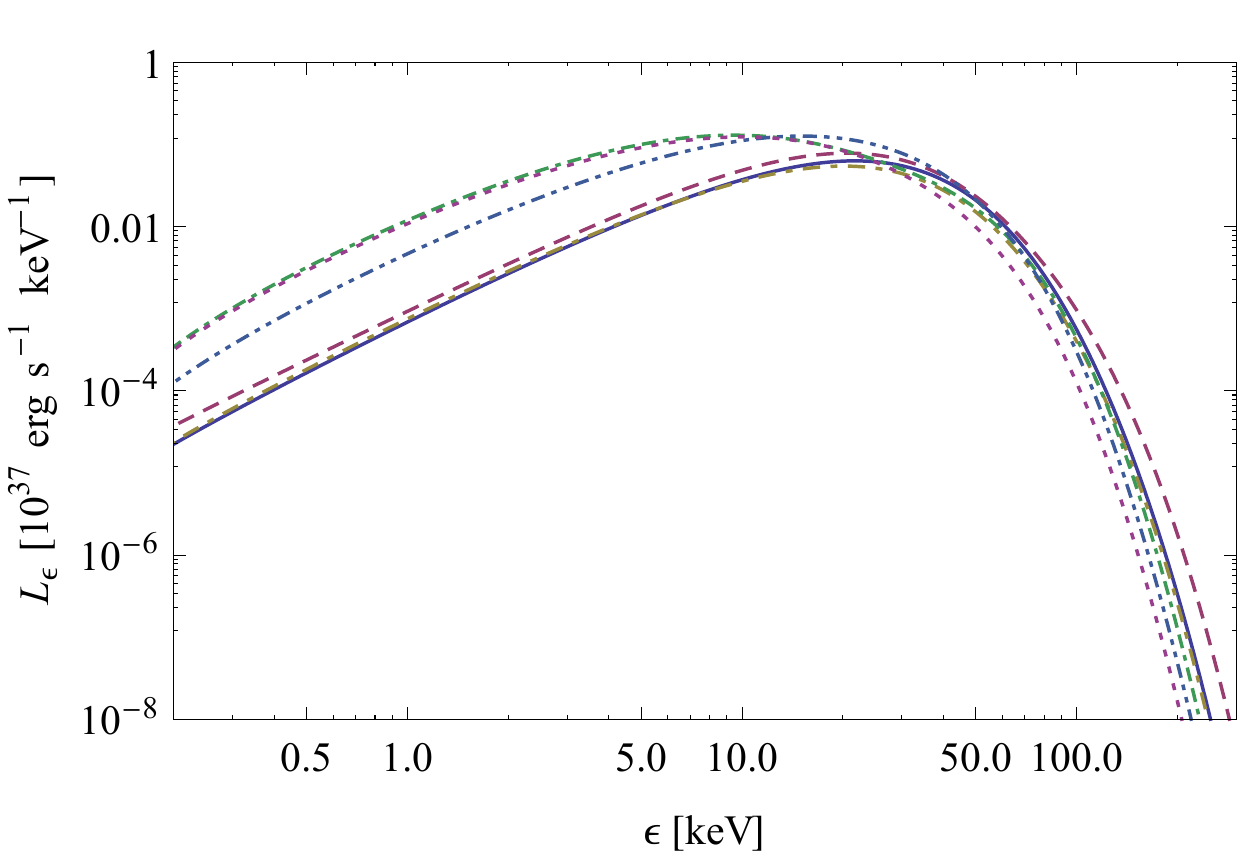}
	\caption{Spectra of radiation from sidewalls of the accretion columns,
calculated simultaneously with velocity and temperature profiles plotted
in Figs \ref{fig:spv} and \ref{fig:dotm2}.
The results correspond to different sets of the model parameters:
`1' (solid line), `2' (dashed), `3' (dot-dashed), `4' (dot-double-dashed), `5' (double-dot-dashed), `6' (dotted).}
  \label{fig:sp}
	\end{center}
 \end{figure}

The resultant velocity distributions
are shown (in terms of quantity $Q$ for uniformity) in left panels of the Fig.~\ref{fig:spv} (for $\dot M_{17}=1$)
and Fig.~\ref{fig:dotm2} (for $\dot M_{17}=2$).
Simultaneously with 2D velocity profiles, the calculations lead to 2D profiles of the Compton temperature (\ref{e:Te})
displayed in  Figs~\ref{fig:spv} and \ref{fig:dotm2} (right panels)
and to the sidewall emergent radiation spectra showed in Fig.~\ref{fig:sp}, where the
column sidewall spectral luminosity
\beq{e:Leps}
L_\epsilon=2\upi r_0\int F_{\epsilon,~r}(r_0, z){\rm d}z
\eeq
is plotted for the different sets of problem parameters.

 The obtained solutions for the velocity of the flow are in a very good conformity with the mound-like 2D shock structures
obtained by solving  the system
including energy equation (\ref{e:energy1}) (i.e., by solving \eq{e:Qeq} in axially symmetrical case).

Each of the Figs \ref{fig:spv} and \ref{fig:dotm2} demonstrates the modifications in
the shape of the front of the shock at fixed accretion rate that occur when the accretion column radius is varied
at fixed $D^\|$
(Figs~\ref{fig:spv}a,~b and Figs~\ref{fig:dotm2}a,~b)
or when $D^\|$ changes at fixed radius
(Figs~\ref{fig:spv}b,~c and Figs~\ref{fig:dotm2}b,~c).

For the convenience, let  each simulation (set of model parameters) be denoted
by the number. At $\dot M_{17}=1$:
`1' $D^{\|}=10D^\perp$, $\tilde{r}_0=6$;
`2' $D^{\|}=10D^\perp$, $\tilde{r}_0=8$;
`3' $D^{\|}=4D^\perp$, $\tilde{r}_0=8$.
At $\dot M_{17}=2$:
`4' and `5' $D^{\|}=10D^\perp$,
`6' $D^{\|}=4D^\perp$;
the column radius is related with $\dot M$ in each case as
\beq{e:raddotM}
\tilde{r}_0=\left.\tilde{r}_0\right|_{\dot M_{17}=1}\dot M_{17}^{1/7}.
\eeq

The temperature (\ref{e:Te})  is close to the electron one established in the area of
the shock mainly by the Compton interaction with radiation at the time-scale \citep{1970JETPL..11...35Z}
\beq{e:tC}
t_{\rm C}\sim \frac{m_{\rm e}c}{\sigma_{\rm T}u}.
\eeq
At a given $r$, the temperature has a minimum located in the settling zone under the shock
(excluding the external radii where, in fact, this zone absents). That is in a qualitative
(and quantitative, in the order of magnitude) agreement with the 1D profiles,
presented by  \cite{2017ApJ...835..129W}, \citep{2017ApJ...835..130W}
(see plots for their $T_{\rm IC}$).
The distributions of the temperature (\ref{e:Te}) depend on the shock wave profile, so
the value $T_{\rm e}$ changes in the radial direction for any fixed $z$ (unless $z$ does not significantly exceed the maximum shock height).
The temperature value corresponding to the bottom boundary (independent of $r$) is dropped in the figures.

In the region of the shock, the time-scale (\ref{e:tC}) is significantly (by several orders of magnitude)
less than both the time of diffusion of photons to the column boundary and the time of energy exchange between
matter and radiation due to free-free processes. Thus, the local Compton equilibrium is a good approximation here.
In the region that is external with respect to the shock,
due to a decrease in the total radiation energy density, $t_{\rm C}$ begins to increase with increasing  $Q$
and becomes comparable with the mentioned time-scales. Therefore, only in this region, where $Q \simeq 1$,
the electron temperature does not attain the values given by expression (\ref{e:Te}).
However, the number of scattering events that photons undergo in this region is
significantly less compared to the number of scatterings experienced in the shock wave, where
 the mean free path of photons is very short.
Thus, one can believe that an overestimation of temperature above the shock
does not significantly distort the effect of thermal Comptonization upon spectrum formation.

The current solutions  (Fig.~\ref{fig:sp}) demonstrate
the dependence of the form of emerging radiation spectra on the magnitudes of velocity divergence and
temperature in the shock.  In the model under consideration, the temperature and the velocity are not independent
and take on the values that provide an appropriate rate of the escape of photons from the column.
It seems that a widening of the shock conditionates the increase of
characteristic bulk Comptonization time-scale $(\nabla\cdot \bm\varv)^{-1}$.
In considering case, at fixed mass accretion rate (and at fixed $D^\perp$ and $D^\|$) this alteration is determined by the accretion
column radius.

The increase of mass accretion rate up to the value $\dot M_{17}=2$ 
leads to a narrowing of the shock and, consequently, to the growth
of the value of the bulk Comptonization rate $\nabla\cdot \bm\varv$.
The characteristic electron temperature slightly decreases with increasing $\dot M$
(compare, for example, Figs~\ref{fig:spv}d and \ref{fig:dotm2}d).
In calculations `3' and `6' (Figs~\ref{fig:spv}f and \ref{fig:dotm2}f), the temperature is practically invariable
near the column axis.

The difference between the spectra at $\dot M_{17}=1$ is not very pronounced, the maxima are at $\epsilon \simeq 20~{\rm keV}$.
The spectra `4' and `6' having the maxima at $\epsilon \simeq 9.5$--$10~{\rm keV}$ remind the upper power-law solutions in fig. 1 of \cite{1982SvAL....8..330L} at high energies ($\epsilon \sim 10$--$50~{\rm keV}$), before the exponential cutoff. Meanwhile, the spectrum `5' has a distinct maximum at $\epsilon \simeq 15~{\rm keV}$.

There is also a qualitative agreement with the solutions of spatially 1D \eq{e:BP} which
have been obtained numerically by \cite{2012A&A...538A..67F} for
the phenomenological
power-law velocity profiles, $\varv\propto z^\eta$, where index $\eta$
is one of the parameters of the problem.
A decrease of $\eta$ (flattening the velocity profile) leads to a hardening
of spectrum tail at fixed column optical depth, electron temperature, and other parameters.
This result was improved by \cite{2016A&A...591A..29F}, where the additional terms have been introduced to
the spatially 1D \eq{e:BP} accounting approximately bremsstrahlung  and sources of the fresh photons
within the column volume.
Nevertheless, the primary part of seed photons is generated under the neutron star surface,  where free-free
mechanisms play the essential role. Conversely, due to relations between the characteristic
time-scales mentioned above, in the current approach, free-free interactions within the column are assumed
not to be  affected significantly on the radiation transfer in the shock wave, where the kinetic energy of the plasma flow is
converted into radiation energy mainly via Compton mechanism.

In order to characterize the shape of the spectra shown in Fig.~\ref{fig:sp}, the soft and hard X-ray colours (hardness ratios)
determined by X-ray luminosity in specific energy bands as (e.g., \citealt{Reig:Nespoli:13})
\beqa{e:HR}
{\rm SC}_1=\frac{L_{7-10~{\rm keV}}}{L_{4-7~{\rm keV}}},~~
{\rm SC}_2=\frac{L_{5-12~{\rm keV}}}{L_{1-3~{\rm keV}}},~~
{\rm HC}=\frac{L_{15-30~{\rm keV}}}{L_{10-15~{\rm keV}}},
\eeqa
have been calculated.  The values are shown in Table~\ref{tab:HR} which demonstrates that the
changes of the model parameters (including mass accretion rate) can lead to the
variations of the hardness of spectrum of the direct column emission. At fixed $D^\|$,
the spectrum becomes softer with increasing $\dot M$.

\begin{figure}
 	\begin{center}
 	\includegraphics[width=0.45\textwidth]{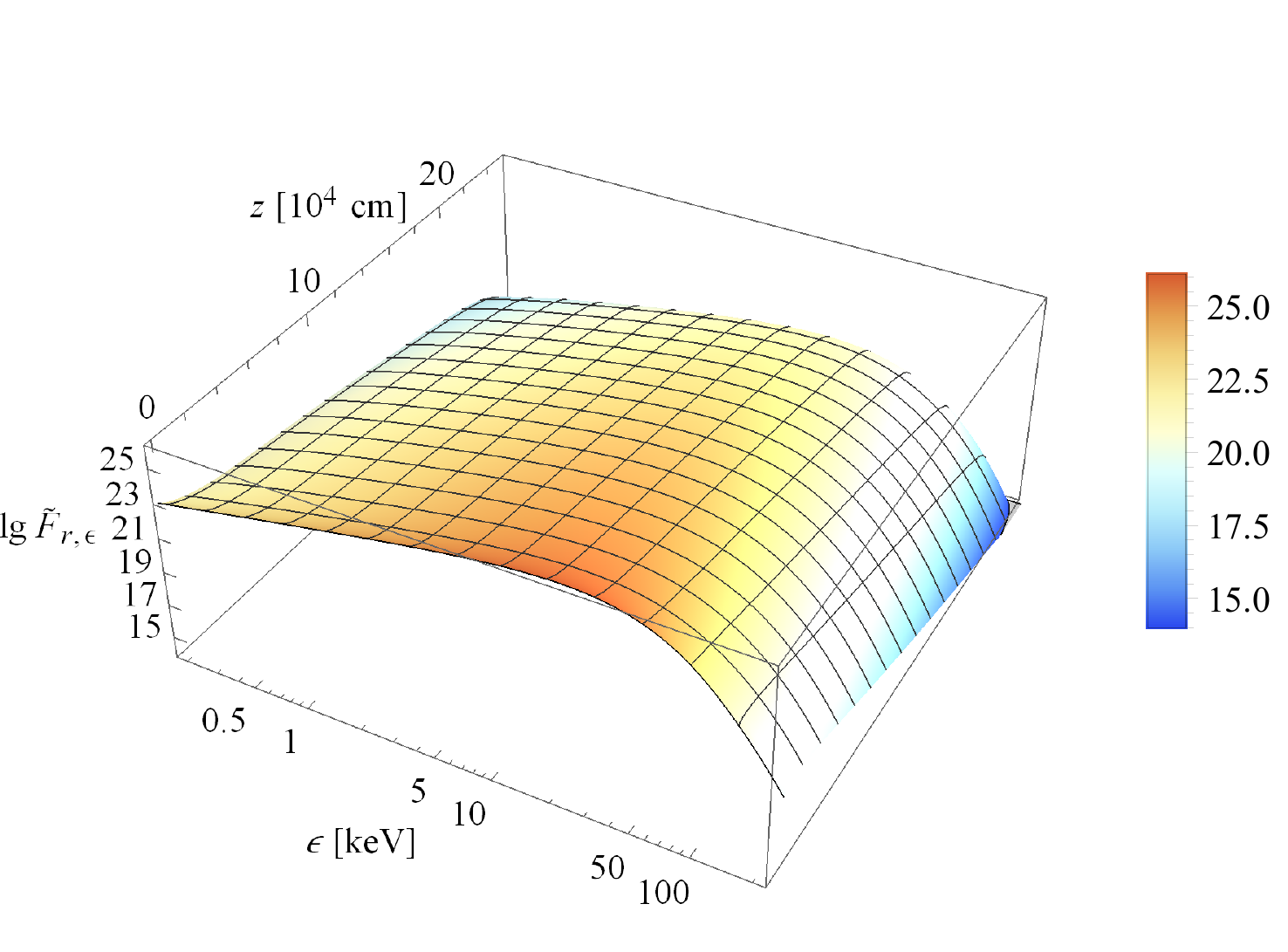}\\(a)\\\vspace{5pt}
  \includegraphics[width=0.45\textwidth]{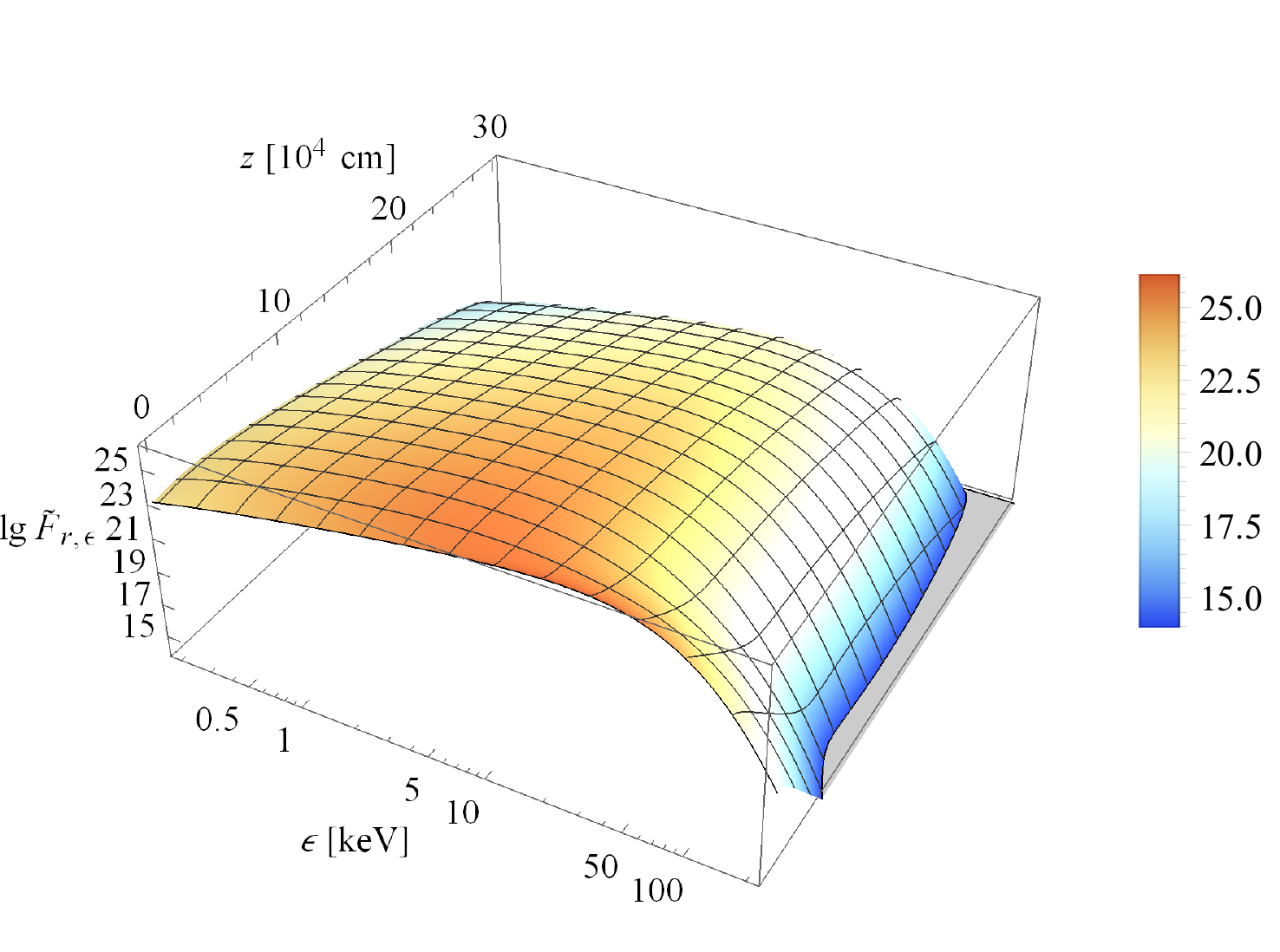}\\(b)
	\caption{Spectral flux from the column sidewall plotted in the dependence on the height.
Here, $\tilde{F}_{\epsilon,~r}={F}_{\epsilon,~r}(r_0,z,\epsilon)/(\rm {erg~ s^{-1} ~cm^{-2}~keV^{-1}})$.
The results correspond  to simulations `1' (a) and `4' (b).}
  \label{fig:Fepsz}
	\end{center}
 \end{figure}

The radial component of spectral radiation flux calculated along the column sidewall, ${F}_{\epsilon,~r}(r_0,z,\epsilon)$, is plotted
in Fig.~\ref{fig:Fepsz} for the cases of $\dot M_{17}=1$ and $\dot M_{17}=2$ corresponding to calculations `1' and `4'.
 The maximal value of this quantity $\sim 10^{26} ~{\rm erg~ s^{-1} ~cm^{-2}~keV^{-1}}$
holds for all simulations `1'--`6'.

\begin{table}
\centering
\caption{X-ray colours of the calculated spectra.}
\label{tab:HR}
\begin{tabular}{cccccc}
\hline
 Spectrum (set of & SC$_1$ & SC$_2$ & HC \\
  model parameters) &  &  &  \\
\hline
   1  & 1.88  & 37.05 & 3.87  \\
   2  & 1.88  & 36.78 & 3.68  \\
   3  &  1.82 & 33.68 & 3.55  \\
   4  & 1.21  & 12.01 & 1.89  \\
   5  & 1.51 & 20.74 & 2.64  \\
   6  & 1.24 & 12.49 & 1.87  \\
\hline
\end{tabular}
\end{table}

\section{Remarks and conclusions}
\label{sec:concl}
The transition to 3D modelling is inevitably accompanied with increasing number of parameters of the problem,
and  only several particular results have been considered above.
The numerical computations of Section \ref{sec:3d} indicate the extent of the distinction of solutions
from axially symmetric models
and lead to the 2D asymmetric distributions of the flux over the surface of the column,
which can be applied to the modelling of the observable pulse profiles
of the emission of X-ray pulsars.

The  height of the shock in the model of narrow unclosed hollow column
 is relatively small compared to the value in the filled-cylinder one.
The radiation leaves the sidewalls
predominantly near the neutron star surface independently of the geometry.
The spectral distribution of this emission is determined by the
 solution of spectral-dependent problem which still contains the set of simplifications
related mainly to using grey diffusion coefficients.
Nevertheless, the results are important for a comparison
of  the column structure, resultant from the frequency-integral
approaches (\citealt{1973NPhS..246....1D}, \citealt{2015MNRAS.452.1601P})
and the results presented in the Section \ref{sec:3dsp}, containing,
 furthermore, 2D distribution of the electron temperature computed in the assumption of the local Compton equilibrium
 and the spectrum calculated in each point of the column.
The conformity of solutions for the velocity signifies  the correspondence of the
obtained spectra to previous 2D shock models.

In the work of \cite{1982SvAL....8..330L} the
shape of spectra was investigated in the dependence on the parameter
related with the value of separability constant and hence with the rate of escape of photons from the column.
Considering the spatially 1D kinetic equation,
the authors are not interested in certain distribution
of the  occupation number over the emitting region surface, which
is necessary for the accurate modelling of the beam of column emission.
In the current work, it is shown that
the equations lead to the exact 2D solution for the structure (within the framework
of the model assumptions) simultaneously with the determination of $T_{\rm e}$ and the modelling
of the spectral radiative transfer, without solving preliminary
problems of determining the structure of the column.

It is clear that changing mass accretion rate may lead to significant modifications of the
form of spectra at high energies $\epsilon>kT_{\rm e}$
and variations in the soft hardness ratios as well.

\begin{figure}
  	\begin{center}
  \includegraphics[width=0.22\textwidth]{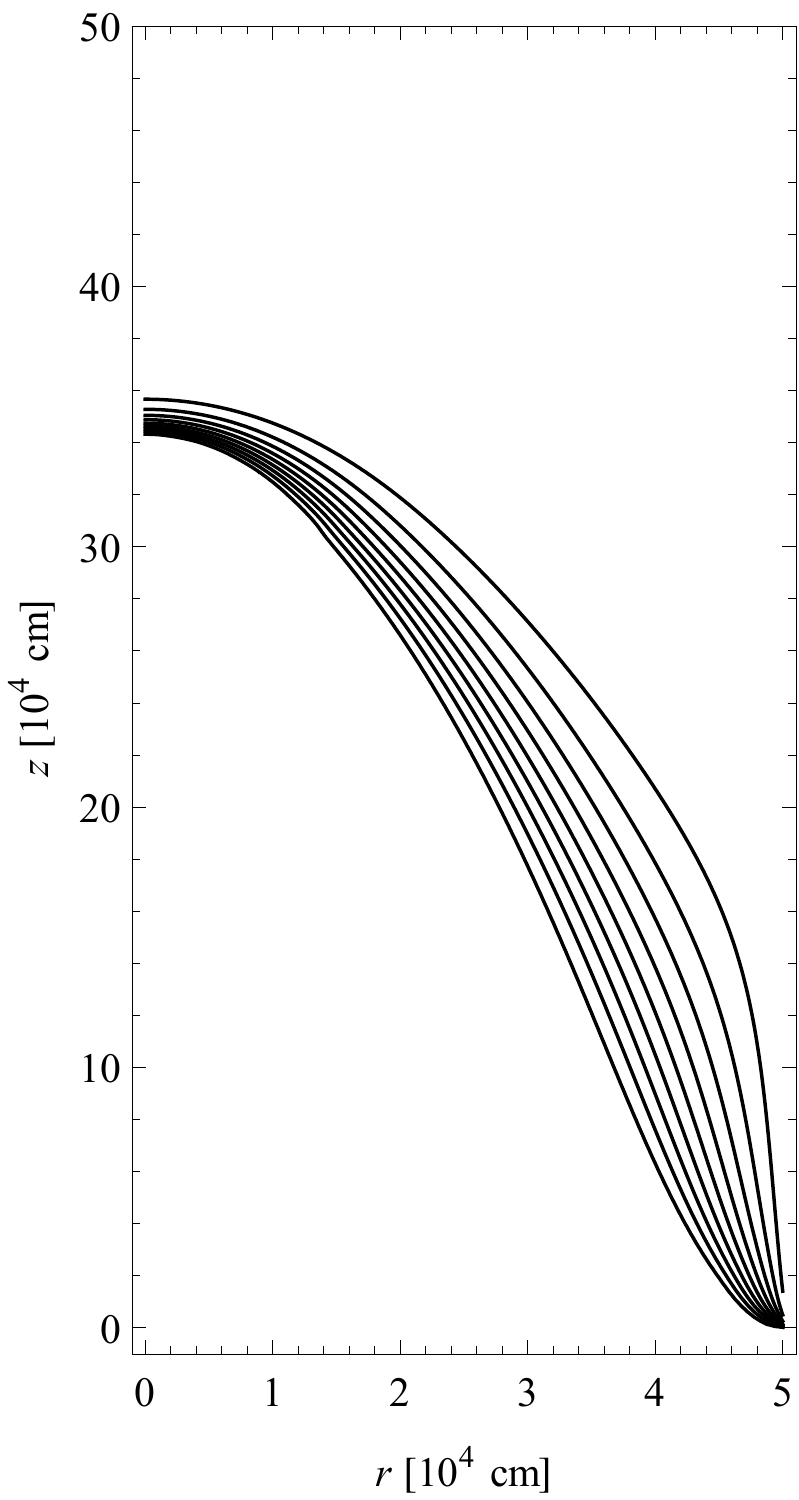}
   \includegraphics[width=0.22\textwidth]{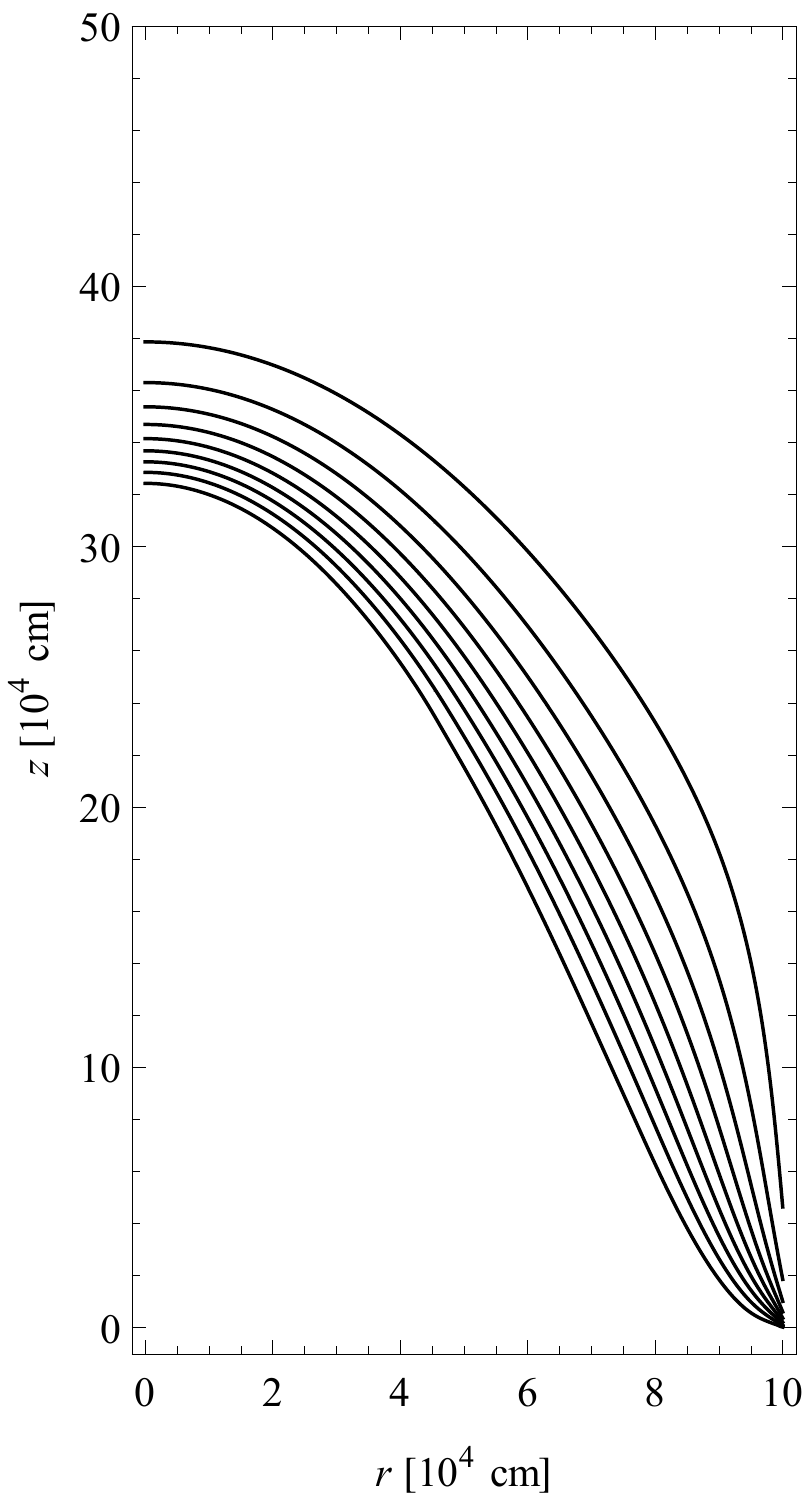}\\(a)\\\vspace{10pt}
    \includegraphics[width=0.15\textwidth]{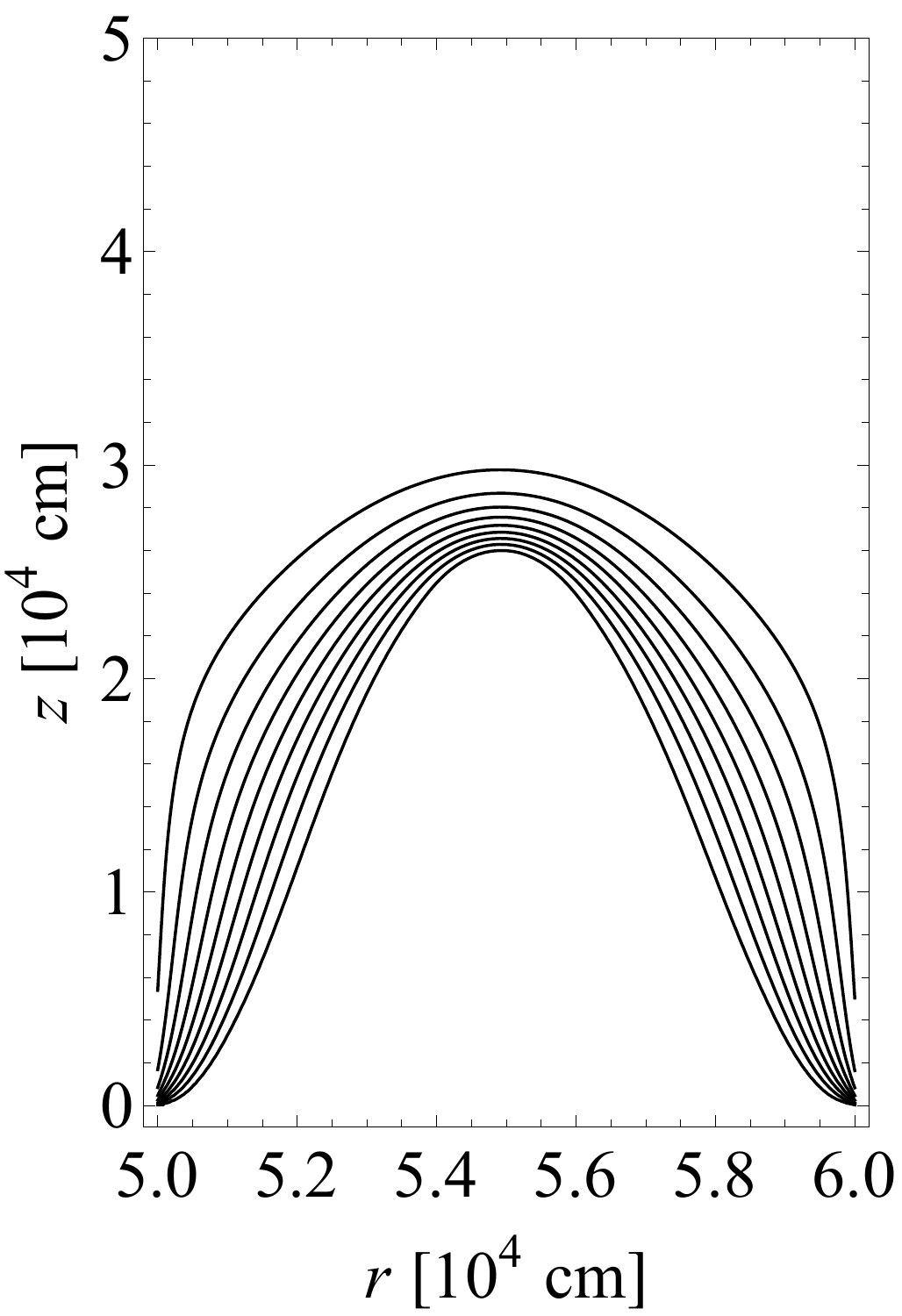}
     \includegraphics[width=0.15\textwidth]{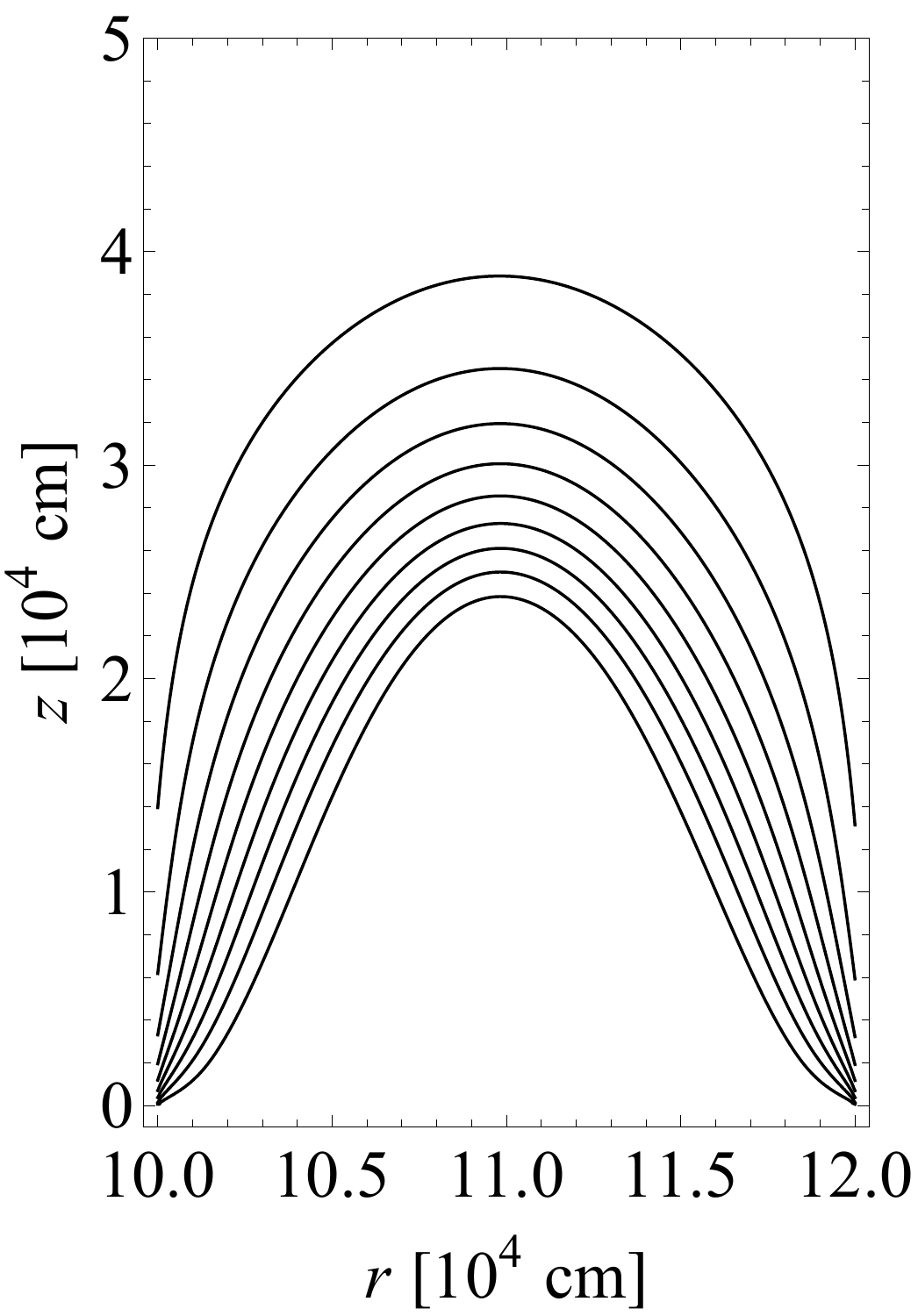}
      \includegraphics[width=0.15\textwidth]{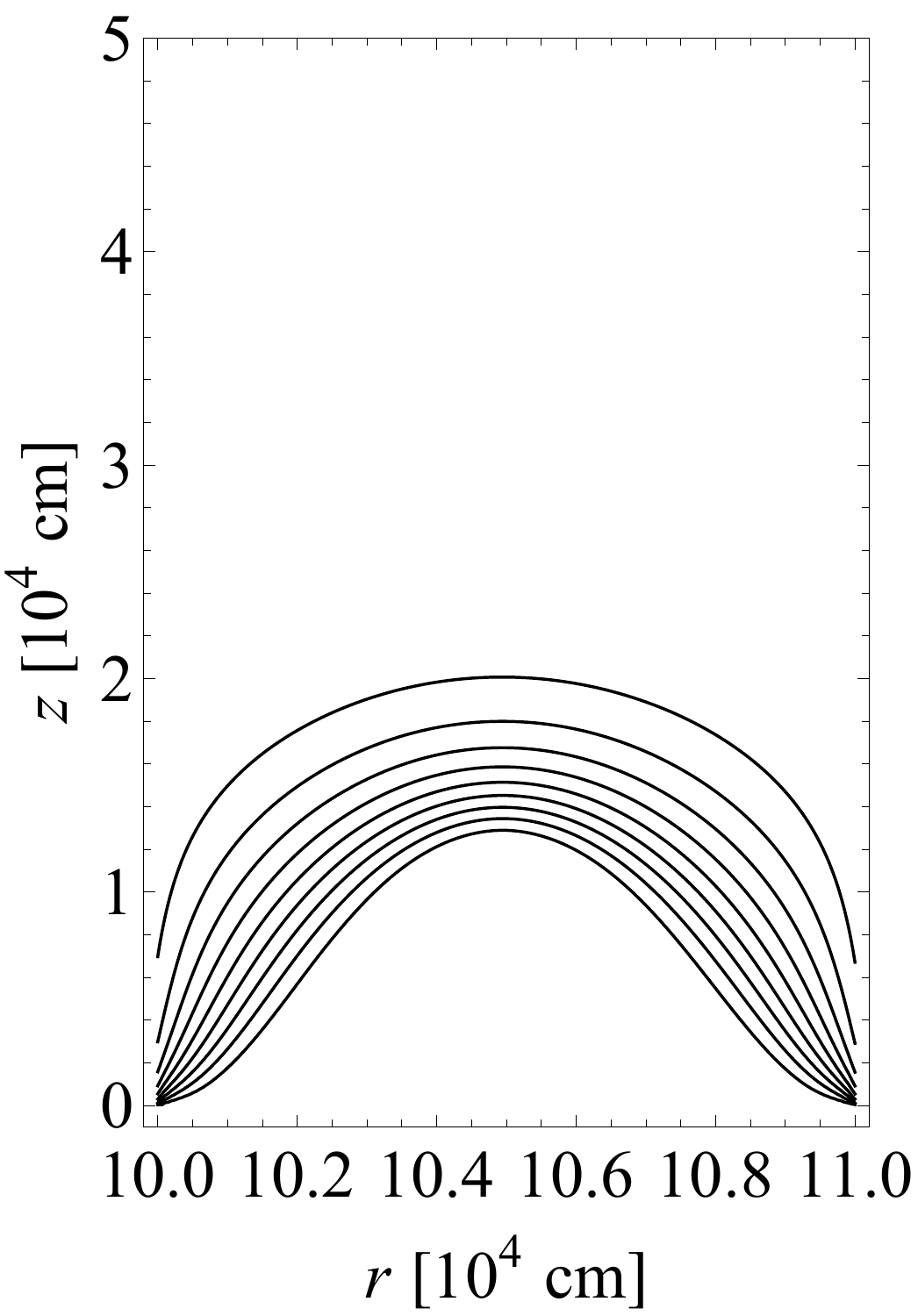}\\(b)\\
	\caption{Contours of equal $Q$ in 2D column models (from 0.9 to 0.1, from top to bottom with interval 0.1). Panels (a) show the solutions for
the filled column at accretion rate $\dot M_{17}=3$: $\tilde{r}_0=5$, left panel and $\tilde{r}_0=10$, right panel.
Panels (b)  show the solutions for the hollow column at accretion rate
$\dot M_{17}=5$. Left panel: $\tilde{r}_0=5$, $b=0.2$; middle panel: $\tilde{r}_0=10$, $b=0.2$;
right panel: $\tilde{r}_0=10$, $b=0.1$.}
  \label{fig:2d}
	\end{center}
 \end{figure}

In the frames of both approaches described in the  current paper (cf. \ref{sec:3d} and \ref{sec:3dsp}) the distribution of dimensionless
velocity $\varv/\varv_0$ (or quantity $Q$) does not practically depend on the
specific value of the velocity  at the upper boundary (see Section \ref{sec:3d}).
One can see that immediately from \eq{e:Qeq} which includes the value of mass flux $S$.

In considering models, the shock height being measured, for example, along the column axis
to the spatial middle of the transitional zone
does not significantly depend  on the
column radius at fixed mass accretion rate. In the case of an axially symmetric ring-like column
transverse section with inner radius $r_0$, the height depends on
the ratio $b$ of thickness of the channel, $br_0$, to $r_0$, but also does not depend on the
specific value of the radius.  In order to show this numerically, and for higher accretion rates than considered
above, let us adduce several solutions for the structure of axially symmetrical columns.
These results are obtained by the
numerical solution of equation (\ref{e:Qeq})
in cylindrical coordinates (2D case). The boundary conditions consist with described in Section \ref{sec:3d}, and at the right
boundaries, within the framework of both models, the condition for the radial component of
radiation flux of type of condition (\ref{e:boundSW}) is set. At the axis of the filled column, $\partial u/\partial r=0$.
Fig.~\ref{fig:2d} indicates the modifications of the shock form for the filled column (a)
and hollow column (b) (the values of $\dot M$ are chosen from the reasons to illustrate all effects distinctly).

The mentioned statements correspond to simple analytic reasoning, described also by  \cite{2015MNRAS.447.1847M},
\cite{2015MNRAS.452.1601P}.
Actually, the equiparation of the radiation diffusion and matter settling times for  the shock of height $H$ indicates
that $H\propto\dot M$ and $H\propto b\dot M$ for the filled and  hollow-cylinder geometry, respectively.
From Fig.~\ref{fig:2d}, however, it follows that the dependence takes place also for the width of the shock
(unaccounted for in the qualitative relations above) which rises with $r_0$,
when the pre-shock density of the flow (at constant $\varv_0$) decreases (see also Figs~\ref{fig:spv}a,~b and \ref{fig:dotm2}a,~b).
The shock width rises, moreover, with
increasing  $D^\|$  (Figs~\ref{fig:spv}b,~c and \ref{fig:dotm2}b,~c).
In all calculations illustrated by Fig.~\ref{fig:2d}, it is set that $D^\|=10D^\perp$.
Since the column radius depends on the magnetic field strength, such a property
may be of interest  in the framework of models implying a connection
between  variations in cyclotron resonance scattering feature energy with $\dot M$
and changes of characteristic height of the shock.

\section*{Acknowledgements}
I thank the anonymous reviewer for useful comments.
This work was supported by RFBR according to the research project 18-32-00890
(development of self-consistent model)
and  the grant 17-15-506-1 of the Foundation for the advancement of theoretical physics and mathematics `BASIS'
(construction of the difference schemes underlying self-consistent modelling and creation of spatially three-dimensional models).

\section*{Data availability}
The calculations described in this paper were performed using the
private codes developed by the author.
The data presented in the figures are available on reasonable request.

\bibliographystyle{mnras}
\bibliography{column}

\begin{thebibliography}{}
\makeatletter
\relax
\def\mn@urlcharsother{\let\do\@makeother \do\$\do\&\do\#\do\^\do\_\do\%\do\~}
\def\mn@doi{\begingroup\mn@urlcharsother \@ifnextchar [ {\mn@doi@}
  {\mn@doi@[]}}
\def\mn@doi@[#1]#2{\def\@tempa{#1}\ifx\@tempa\@empty \href
  {http://dx.doi.org/#2} {doi:#2}\else \href {http://dx.doi.org/#2} {#1}\fi
  \endgroup}
\def\mn@eprint#1#2{\mn@eprint@#1:#2::\@nil}
\def\mn@eprint@arXiv#1{\href {http://arxiv.org/abs/#1} {{\tt arXiv:#1}}}
\def\mn@eprint@dblp#1{\href {http://dblp.uni-trier.de/rec/bibtex/#1.xml}
  {dblp:#1}}
\def\mn@eprint@#1:#2:#3:#4\@nil{\def\@tempa {#1}\def\@tempb {#2}\def\@tempc
  {#3}\ifx \@tempc \@empty \let \@tempc \@tempb \let \@tempb \@tempa \fi \ifx
  \@tempb \@empty \def\@tempb {arXiv}\fi \@ifundefined
  {mn@eprint@\@tempb}{\@tempb:\@tempc}{\expandafter \expandafter \csname
  mn@eprint@\@tempb\endcsname \expandafter{\@tempc}}}

\bibitem[\protect\citeauthoryear{{Basko} \& {Sunyaev}}{{Basko} \&
  {Sunyaev}}{1976}]{1976MNRAS.175..395B}
{Basko} M.~M.,  {Sunyaev} R.~A.,  1976, \mn@doi [\mnras]
  {10.1093/mnras/175.2.395}, \href
  {http://adsabs.harvard.edu/abs/1976MNRAS.175..395B} {175, 395}

\bibitem[\protect\citeauthoryear{{Becker} \& {Wolff}}{{Becker} \&
  {Wolff}}{2007}]{2007ApJ...654..435B}
{Becker} P.~A.,  {Wolff} M.~T.,  2007, \mn@doi [\apj] {10.1086/509108}, \href
  {http://adsabs.harvard.edu/abs/2007ApJ...654..435B} {654, 435}

\bibitem[\protect\citeauthoryear{{Blandford} \& {Payne}}{{Blandford} \&
  {Payne}}{1981}]{1981MNRAS.194.1033B}
{Blandford} R.~D.,  {Payne} D.~G.,  1981, \mn@doi [\mnras]
  {10.1093/mnras/194.4.1033}, \href
  {http://adsabs.harvard.edu/abs/1981MNRAS.194.1033B} {194, 1033}

\bibitem[\protect\citeauthoryear{{Canuto}, {Lodenquai}  \& {Ruderman}}{{Canuto}
  et~al.}{1971}]{1971PhRvD...3.2303C}
{Canuto} V.,  {Lodenquai} J.,   {Ruderman} M.,  1971, \mn@doi [\prd]
  {10.1103/PhysRevD.3.2303}, \href
  {https://ui.adsabs.harvard.edu/abs/1971PhRvD...3.2303C} {3, 2303}

\bibitem[\protect\citeauthoryear{{Davidson}}{{Davidson}}{1973}]{1973NPhS..246....1D}
{Davidson} K.,  1973, \mn@doi [Nature Physical Science]
  {10.1038/physci246001a0}, \href
  {http://adsabs.harvard.edu/abs/1973NPhS..246....1D} {246, 1}

\bibitem[\protect\citeauthoryear{{Farinelli}, {Ceccobello}, {Romano}  \&
  {Titarchuk}}{{Farinelli} et~al.}{2012}]{2012A&A...538A..67F}
{Farinelli} R.,  {Ceccobello} C.,  {Romano} P.,   {Titarchuk} L.,  2012,
  \mn@doi [\aap] {10.1051/0004-6361/201118008}, \href
  {http://adsabs.harvard.edu/abs/2012A%26A...538A..67F} {538, A67}

\bibitem[\protect\citeauthoryear{{Farinelli}, {Ferrigno}, {Bozzo}  \&
  {Becker}}{{Farinelli} et~al.}{2016}]{2016A&A...591A..29F}
{Farinelli} R.,  {Ferrigno} C.,  {Bozzo} E.,   {Becker} P.~A.,  2016, \mn@doi
  [\aap] {10.1051/0004-6361/201527257}, \href
  {http://adsabs.harvard.edu/abs/2016A%26A...591A..29F} {591, A29}

\bibitem[\protect\citeauthoryear{{Gnedin} \& {Nagel}}{{Gnedin} \&
  {Nagel}}{1984}]{1984A&A...138..356G}
{Gnedin} I.~N.,  {Nagel} W.,  1984, \aap, \href
  {http://adsabs.harvard.edu/abs/1984A%26A...138..356G} {138, 356}

\bibitem[\protect\citeauthoryear{{Kompaneets}}{{Kompaneets}}{1956}]{1956Kompaneets}
{Kompaneets} A.~S.,  1956, Zh. Eksp. Teor. Fiz., 31, 876

\bibitem[\protect\citeauthoryear{{Lodenquai}, {Canuto}, {Ruderman}  \&
  {Tsuruta}}{{Lodenquai} et~al.}{1974}]{1974ApJ...190..141L}
{Lodenquai} J.,  {Canuto} V.,  {Ruderman} M.,   {Tsuruta} S.,  1974, \mn@doi
  [\apj] {10.1086/152858}, \href
  {https://ui.adsabs.harvard.edu/abs/1974ApJ...190..141L} {190, 141}

\bibitem[\protect\citeauthoryear{{Lyubarskii} \& {Syunyaev}}{{Lyubarskii} \&
  {Syunyaev}}{1982}]{1982SvAL....8..330L}
{Lyubarskii} Y.~E.,  {Syunyaev} R.~A.,  1982, Soviet Astronomy Letters, \href
  {https://ui.adsabs.harvard.edu/abs/1982SvAL....8..330L} {8, 330}

\bibitem[\protect\citeauthoryear{{Meszaros}}{{Meszaros}}{1984}]{1984SSRv...38..325M}
{Meszaros} P.,  1984, \mn@doi [\ssr] {10.1007/BF00176833}, \href
  {http://adsabs.harvard.edu/abs/1984SSRv...38..325M} {38, 325}

\bibitem[\protect\citeauthoryear{{Mushtukov}, {Suleimanov}, {Tsygankov}  \&
  {Poutanen}}{{Mushtukov} et~al.}{2015}]{2015MNRAS.447.1847M}
{Mushtukov} A.~A.,  {Suleimanov} V.~F.,  {Tsygankov} S.~S.,   {Poutanen} J.,
  2015, \mn@doi [\mnras] {10.1093/mnras/stu2484}, \href
  {https://ui.adsabs.harvard.edu/abs/2015MNRAS.447.1847M} {447, 1847}

\bibitem[\protect\citeauthoryear{{Postnov}, {Gornostaev}, {Klochkov},
  {Laplace}, {Lukin}  \& {Shakura}}{{Postnov}
  et~al.}{2015}]{2015MNRAS.452.1601P}
{Postnov} K.~A.,  {Gornostaev} M.~I.,  {Klochkov} D.,  {Laplace} E.,  {Lukin}
  V.~V.,   {Shakura} N.~I.,  2015, \mn@doi [\mnras] {10.1093/mnras/stv1393},
  \href {http://adsabs.harvard.edu/abs/2015MNRAS.452.1601P} {452, 1601}

\bibitem[\protect\citeauthoryear{{Reig} \& {Nespoli}}{{Reig} \&
  {Nespoli}}{2013}]{Reig:Nespoli:13}
{Reig} P.,  {Nespoli} E.,  2013, \aap, 551, A1

\bibitem[\protect\citeauthoryear{{Samarskii}}{{Samarskii}}{1962}]{1962samarskii}
{Samarskii} A.~A.,  1962, \mn@doi [Zh. Vychisl. Mat. Mat. Fiz.]
  {10.1051/0004-6361/201834479}, \href
  {https://ui.adsabs.harvard.edu/abs/2019A&A...622A..61S} {2, 25}

\bibitem[\protect\citeauthoryear{{Samarskii, A.~A.}}{{Samarskii,
  A.~A.}}{2001}]{2001samarskii_eng}
{Samarskii, A.~A.} 2001, The theory of difference schemes.
Marcel Dekker, inc., New York, Base

\bibitem[\protect\citeauthoryear{{Wang} \& {Frank}}{{Wang} \&
  {Frank}}{1981}]{1981A&A....93..255W}
{Wang} Y.-M.,  {Frank} J.,  1981, \aap, \href
  {http://adsabs.harvard.edu/abs/1981A%26A....93..255W} {93, 255}

\bibitem[\protect\citeauthoryear{{West}, {Wolfram}  \& {Becker}}{{West}
  et~al.}{2017a}]{2017ApJ...835..129W}
{West} B.~F.,  {Wolfram} K.~D.,   {Becker} P.~A.,  2017a, \mn@doi [\apj]
  {10.3847/1538-4357/835/2/129}, \href
  {https://ui.adsabs.harvard.edu/abs/2017ApJ...835..129W} {835, 129}

\bibitem[\protect\citeauthoryear{{West}, {Wolfram}  \& {Becker}}{{West}
  et~al.}{2017b}]{2017ApJ...835..130W}
{West} B.~F.,  {Wolfram} K.~D.,   {Becker} P.~A.,  2017b, \mn@doi [\apj]
  {10.3847/1538-4357/835/2/130}, \href
  {https://ui.adsabs.harvard.edu/abs/2017ApJ...835..130W} {835, 130}

\bibitem[\protect\citeauthoryear{{Zel'dovich} \& {Levich}}{{Zel'dovich} \&
  {Levich}}{1970}]{1970JETPL..11...35Z}
{Zel'dovich} Y.~B.,  {Levich} E.~V.,  1970, Soviet Journal of Experimental and
  Theoretical Physics Letters, \href
  {https://ui.adsabs.harvard.edu/abs/1970JETPL..11...35Z} {11, 35}

\makeatother
\end{thebibliography}

\label{lastpage}

\end{document}